\documentclass[review]{elsarticle}
\usepackage{graphicx} 

\usepackage{hyperref}
\hypersetup{
    colorlinks,
    citecolor=blue,
    filecolor=black,
    linkcolor=blue,
    urlcolor=black
}

\usepackage{textcomp}
\usepackage{pgfplots}

\pgfplotsset{compat=1.16} 
\usepackage{import}
\usepackage{pifont}
\newcommand{\cmark}{\ding{51}}%
\newcommand{\xmark}{\ding{53}}%
\usepackage[center]{caption}
\usepackage{multirow}
\usepackage[colorinlistoftodos]{todonotes}
\usepackage{url}
\usepackage{mathtools}
\usepackage[margin=1in]{geometry}

\usepackage{amsmath,amssymb}
\usepackage{soul}

\usepackage{tikz}
\usepackage{forest}
\usetikzlibrary{trees,positioning,shapes,shadows,arrows.meta}
\usepackage{listings}
\usepackage{xcolor}

\begin{document}
\begin{frontmatter}
\title{
A Comprehensive Review of Handling Missing Data: Exploring Special Missing Mechanisms}
    \author[deakin]{Youran Zhou\corref{cor1}}
    \cortext[cor1]{Corresponding author}
    \ead{echo.zhou@deakin.edu.au} 
    \author[deakin]{Sunil Aryal}
    \ead{sunil.aryal@deakin.edu.au} 
    \author[deakin]{Mohamed Reda Bouadjenek}
    \ead{reda.bouadjenek@deakin.edu.au} 
    
    \address[deakin]{School of Information Technology, Deakin University, Waurn Ponds Campus, Geelong, VIC 3216, Australia}

    \begin{abstract}
        Missing data poses a significant challenge in data science, affecting decision-making processes and outcomes. Understanding what missing data is, how it occurs, and why it is crucial to handle it appropriately is paramount when working with real-world data, especially in tabular data, one of the most commonly used data types in the real world. Three missing mechanisms are defined in the literature: Missing Completely At Random (MCAR), Missing At Random (MAR), and Missing Not At Random (MNAR), each presenting unique challenges in imputation. Most existing work are focused on MCAR that is relatively easy to handle. The special missing mechanisms of MNAR and MAR are less explored and understood. This article reviews existing literature on handling missing values. It compares and contrasts existing methods in terms of their ability to handle different missing mechanisms and data types. It identifies research gap in the existing literature and lays out potential directions for future research in the field. The information in this review will help data analysts and researchers to adopt and promote good practices for handling missing data in real-world problems.
    \end{abstract}
\end{frontmatter}

\section{Introduction}

Missing data refers to the absence of values or information in specific fields or attributes within a dataset. In other words, it occurs when data points are unavailable or have not been recorded for certain variables or observations. The presence of missing data can arise for various reasons during data collection, storage, or processing~\cite{1}. For instance, participants might choose not to answer specific questions in surveys or questionnaires, leading to missing data for those specific items. Similarly, in sensor data collected from scientific experiments, missing data can occur if sensors malfunction or fail to record data accurately.\\\\
Handling missing data is of utmost importance in data science, as it presents a significant challenge with the potential to affect decision-making processes and research outcomes adversely. When working with real-world data, dealing with missing values becomes one of the many obstacles encountered. Data analysis tasks' accuracy and efficiency rely heavily on data quality, making it crucial for data analysts and researchers working with tabular data to address the issue of missing values effectively. Mishandling missing values can lead to biased conclusions, compromised generalizability of findings, and hindered the development of robust models. Therefore, exploring and implementing appropriate strategies for handling missing values is essential to ensure accurate and reliable results in data-driven analyses and decision-making processes.
\\\\
The occurrence of missing data can be attributed to various mechanisms, each associated with specific assumptions about the data. Among the commonly recognized mechanisms are Missing Completely At Random (MCAR), Missing At Random (MAR), and Missing Not At Random (MNAR). While each mechanism poses challenges, special attention needs to be given to MAR and MNAR, as they represent the most intricate and least understood scenarios without explicit assumptions or definitions. These mechanisms, which we refer to as \textbf{Special Missing Mechanisms}, are particularly significant due to their association with sensitive data privacy concerns and the potential provision of hidden information. They introduce additional complexity and challenges in accurately imputing missing values, with significant implications for data analysis and decision-making processes. As such, exploring relevant techniques for managing these diverse missing mechanisms becomes even more imperative.
\\\\
The existing literature contains a wealth of studies on missing data imputation. Little \& Rubin~\cite{little} have extensively studied the prevalence of incomplete data, emphasizing the need for robust statistical approaches to address this issue effectively. Several review articles, surveys, and books have been published on the missing data topic. For instance, \cite{healthcareDL} has discussed missing data imputation approaches in healthcare. Norazian et al. \cite{medical2013roles} focus specifically on imputation methods and software for handling missing data in time-series datasets. Garcia et al. \cite{garcia2010pattern} analyze the missing data problem in pattern classification tasks, comparing different methods, including ensemble and fuzzy approaches. Other studies, such as Velasco et al. \cite{velasco2020real}, delve into missing values in sensor data of marine systems. Emmanuel et al. \cite{emmanuel2021survey} and Jegadeeswari et al. \cite{jegadeeswari2022missing} discuss machine learning approaches for handling missing data but do not specifically explore deep learning methods. Adhikari et al. \cite{adhikari2022comprehensive} focus on missing data in IoT, though special missing mechanisms are not explicitly addressed. Liu et al. and Sun et al.  Ma et al. \cite{ma2018bayesian} proposed a Bayesian-based imputation method survey paper, while Graham et al. \cite{graham2009missing} studied the utilization of normal-model multiple imputation. Dong et al. \cite{dong2013principled} compared multiple imputations, full information maximum likelihood, and expectation-maximization algorithms. Sun et al. \cite{dlreview} conducted a review and comparative study on deep learning and traditional machine learning methods. There are also some reviews that focus on other areas \cite{pigott2001review,  baraldi2010introduction, lin2020missing, 10225306}
\\\\However, a notable disparity exists as most of these studies primarily focus on the most common case of MCAR, while fewer delve into the more complex cases of MAR and MNAR. Figure \ref{fig:MAR_MNAR_Article} shows that there are only a very limited numbers of approaches for handling missing data with special missing mechanisms.  Additionally, even for those approaches dealing with special missing mechanisms, a lack of standardized approaches for generating missing data in different experiments hinders our ability to make meaningful comparisons between methods. Consequently, there is a compelling need for a comprehensive survey that accounts for various missing mechanisms, specifically for tabular data --- the most prevalent data type.
\subsection{Contribution}
Our study makes several significant contributions to this field:
\begin{enumerate}
\item  Comprehensive Review of Special Missing Mechanisms in tabular data: We provide a comprehensive summary and in-depth discussion of various methods for handling missing data, particularly focusing on special missing mechanisms in tabular data. Our review covers traditional techniques such as deletion and imputation, as well as emerging methods based on representation learning. We mainly focus on imputation based methods, as modern data sets grow in size and complexity, conventional statistical and machine learning-based approaches may prove insufficient. By emphasizing deep learning-based strategies, our work aims to equip researchers and practitioners with a valuable resource for effectively addressing missing data challenges.
\item Thorough Examination of Missing Data Generation Methods: In our review, we meticulously catalog the different methods used in the generation of missing data, especially for the less frequently addressed Missing at Random (MAR) and Missing Not at Random (MNAR) mechanisms. While prior research primarily focuses on MCAR, we recognize the limited attention given to MAR and MNAR. Our goal is to raise awareness of the importance and variability of special missing mechanisms and encourage a more comprehensive exploration of these mechanisms in future studies.
\item Guidance for Future Research Directions: To further advance the field of imputation techniques, we propose future research directions aimed at overcoming the limitations of existing methods and promoting the adoption of advanced techniques in practical settings. By identifying research gaps within the literature and suggesting new applications for imputation schemes, our study serves as a road map for researchers and practitioners. We aim to facilitate the implementation of imputation approaches across different types of data, ultimately contributing to the advancement of the missing data handling field.
\end{enumerate}
 The rest of this paper is organized as follows: Section \ref{sec:Background and Preliminary} provides the background on the key features of missing data including missing data, missing pattern and missing mechanism, and the common method of handling missing data. 
Section \ref{sec:Existing Methods of Handling Missing Data}, introduce on the taxonomy of handling missing data  techniques.   Section \ref{sec:Deletion for Handling Missing Data}, \ref{sec:imputation methods}, \ref{sec:Representation Learning for Handling Missing Data} mainly introduced the methods that dealing with missing data. Section \ref{sec:Methodology for Missing data Generation} list commonly used missing data generation for Special Missing Mechanism from the literature. Section \ref{sec:Evaluation Metrics} reviews the evaluation metrics used to measure their performance. Furthermore, Section \ref{sec:Limitations} presents challenges and future direction of the works.

\pgfplotsset{/pgf/number format/1000 sep=}

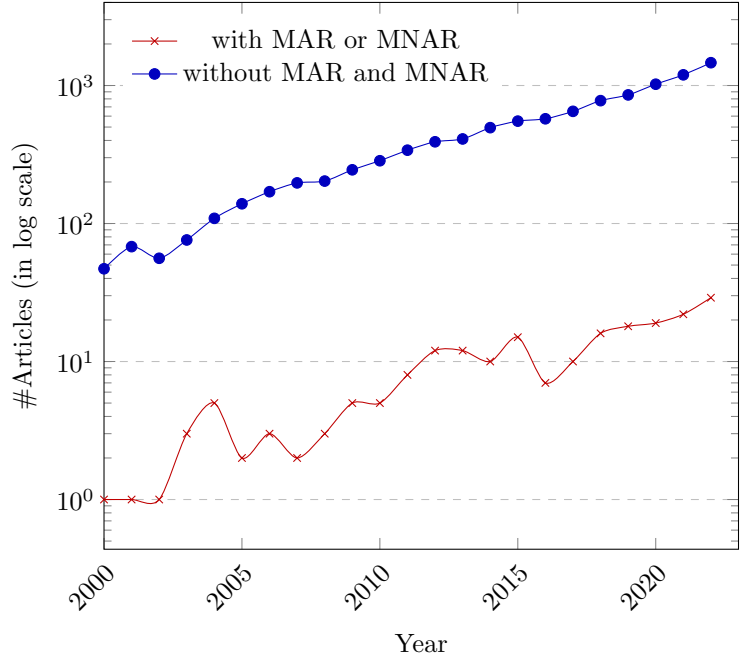
\begin{figure}[!h]
    \centering

\begin{tikzpicture}[yscale=1, xscale = 1]
    \pgfplotsset{
        scale only axis,
        y axis style/.style={
            yticklabel style=#1,
            ylabel style=#1,
            y axis line style=#1,
            ytick style=#1
       }
    }

\begin{axis}[
  ymin=0, ymax=4000,
  ymode=log,
  xmin=2000, xmax = 2023,
  xlabel=Year,
  ylabel= \#Articles (in log scale),
  legend style={draw=none},
  xtick = {2000,2005,2010,2015,2020},
    xticklabel style={rotate=45, anchor=north east},
    legend pos=north west,
    ymajorgrids=true,
    grid style=dashed
]

\addplot[smooth,mark=x,red!75!black]
  coordinates{
(2022,	29)
(2021,	22)
(2020,	19)
(2019,	18)
(2018,	16)
(2017,	10)
(2016,	7)
(2015,	15)
(2014,  10)
(2013,	12)
(2012,	12)
(2011,	8)
(2010,	5)
(2009,	5)
(2008,	3)
(2007,	2)
(2006,	3)
(2005,	2)
(2004,	5)
(2003,	3)
(2002,1)
(2001,1)
(2000,1)
}; \label{plot_one}
\addlegendentry{with MAR or MNAR}
\addplot[smooth,mark=*,blue!75!black]
  coordinates{
(2022,	1463)
(2021,	1195)
(2020,	1022)
(2019,	855)
(2018,	777)
(2017,	650)
(2016,	574)
(2015,	552)
(2014,  495)
(2013,	410)
(2012,	391)
(2011,	340)
(2010,	285)
(2009,	245)
(2008,	203)
(2007,	197)
(2006,	170)
(2005,	139)
(2004,	109)
(2003,	76)
(2002,56)
(2001,68)
(2000,47)
}; \label{plot_one}
\addlegendentry{without MAR and MNAR}
\end{axis}

\end{tikzpicture}

    \caption{Number of articles on handling missing data retrieved by Scopus based on keyword searches with and without the special mechanisms keywords of MAR or MNAR }
    \label{fig:MAR_MNAR_Article}

\end{figure}

\section{Background and Preliminary}
\label{sec:Background and Preliminary}
Let us consider a complete data matrix with with $k$ variables and $n$ instances, $$\boldsymbol{X} = (\boldsymbol{x}_1, ..., \boldsymbol{x}_n)^T \in \mathcal{X}^k$$ In the context of missing data, each sample can be divided into two parts: the observed part and the missing part denoted as $X = (X^o, X^m)$. Here, $X^o$ represents the part without any missing values, while $X^m$ contains the missing values. There is another $n \times k $ matrix of missing data indicators (Mask) for $\boldsymbol{X}$ is denoted as $\boldsymbol{M}$, where an element $m_{ij}$ for individual i on variable j is $\boldsymbol{M_{ij}} = 0$  for missing values and $\boldsymbol{M_{ij}} = 1$ for observed values. The symbol $\Psi$ denotes the \textit{missing parameters}. The \textit{missing parameters} describes how missing values are generated with respect to the conditional distribution of missing. This distribution can be expressed as:
$$f(\boldsymbol{M} |  \boldsymbol{X}, \Psi)$$ 
That means the conditional distribution of Mask $\boldsymbol{M}$ depends on the data values in $\boldsymbol{X}$ and the missing parameters $\Psi$. There are three aspects that controlled by the missing parameters $\Psi$: Missing Rate, Missing Pattern and Missing Mechanisms.

\subsection{Missing Rate}
The missing rate refers to the proportion of missing data in the entire dataset. It plays a crucial role in determining the difficulty of solving the missing data problem. A higher missing rate indicates that a larger portion of the data is missing, resulting in less information available for algorithms to analyze, thereby making the task more challenging. However, the impact of missing data on algorithms depends on the problem setting and data distribution. In some cases, missing data may consist of outliers or provide less important information, resulting in minimal impact on the algorithms' performance.
\begin{figure}[!h]
    \centering
    \includegraphics[width=0.9\linewidth]{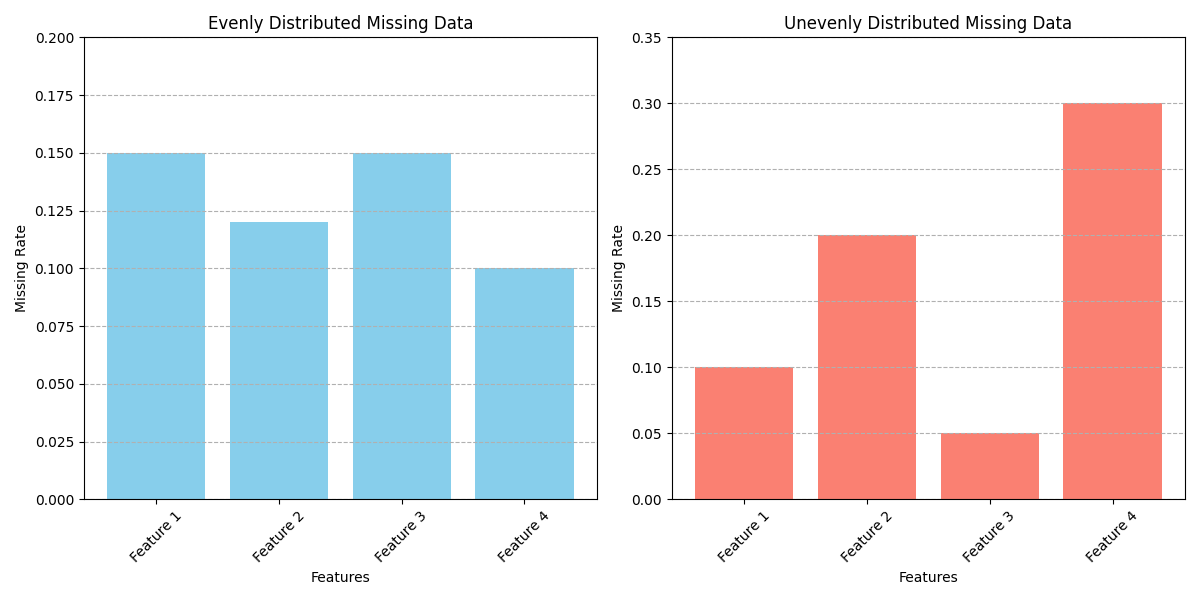}
    \caption{Comparison of missing rates in different features. The left plot illustrates evenly distributed missing data, where each feature has similar missing rates. On the right plot, unevenly distributed missing data is shown, with varying missing rates across features. This visual representation highlights the diversity in missing data patterns within the dataset.}
    \label{fig:missing rate}
\end{figure}
\\\\The missing rate can be evenly distributed or unevenly distributed across the dataset. For example, in tabular datasets, missing data can occur in multiple columns, where each column may have the same missing rate, or some columns may have a higher missing rate than others (See Figure\ref{fig:missing rate}). In image datasets, some images may have a few missing pixels randomly occur, while others may have a significant part of the image missing. Similarly, in multimodality sensor datasets, some sensors may have been malfunctioning for an extended period, leading to a large amount of missing information, while other sensors may have only lost one or two signal readings. The severity of missing data can have a significant impact on the performance of models and analyses. Therefore, understanding the missing rate and its implications is also a crucial aspect of research and analysis.

\subsection{Missing Pattern}

    \begin{figure}[!h]
    \centering
    \includegraphics[width=1\linewidth]{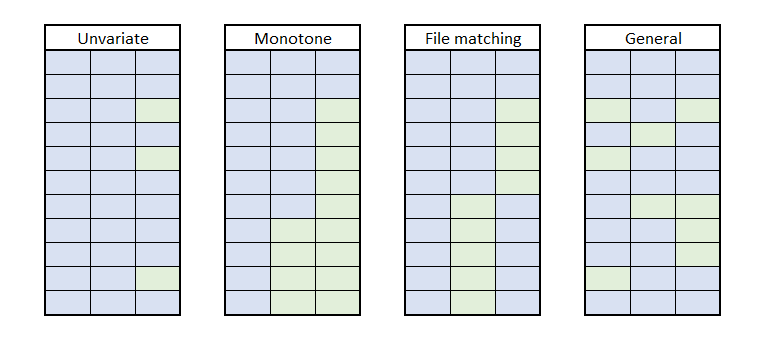}
    \caption{Missing data patterns in multivariate data. Blue is observed  $X^o$, green is missing part $X^m$. The missing pattern from left to right are: \textit{Univariate, Monotone, Filematching and General}}
    \label{fig:Missing Pattern}
\end{figure}
In the existing literature~\cite{missingpattern}, it has been observed that missing patterns frequently occur in multivariate datasets due to various reasons (see Figure \ref{fig:Missing Pattern}).
These missing patterns in multivariate data can be categorized into different types:\\\\
\textbf{Univariate and Multivariate Patterns:} \\In the Univariate pattern, missing values occur in only one variable or column of the dataset. For instance, in a dataset containing information about age, income, and education, the univariate missing pattern occurs when one of these variables has missing values, while the others have complete data.(see Figure \ref{fig:Missing Pattern} \textit{Univariate})\\\\
\textbf{Monotone and Non-Monotone Patterns:}\\
 In a monotone missing pattern, missing values occur systematically in one direction within the dataset. This implies that once a value is missing, all subsequent values in the dataset are also missing. Monotone missing patterns can be classified into "Monotone Missing Completely At Random (MCAR)" or "Monotone Missing Not At Random (MNAR)". (see Figure \ref{fig:Missing Pattern} \textit{Monotone})\\\\
\textbf{Connecting and Unconnecting Patterns:}\\
The figure \ref{fig:Missing Pattern} \textit{File matching} depicts a connecting pattern, where it is possible to travel to all blue cells by horizontal or vertical moves. However, this pattern becomes unconnected if we remove the first column.\\\\
The issue of missing patterns is pervasive and occurs across various data types, such as tabular data, time series data, image data, sensor data, survey data, and biomedical data. While these missing patterns present challenges in data analysis and modeling, our review focuses specifically on tabular datasets due to their prevalence and significance in diverse fields.  While recognizing the complexity of missing patterns in other data types, we prioritize a detailed investigation of missing mechanisms in tabular data rather than missing pattern.

\subsection{Missing Mechanisms}

\begin{table}[ht]
\centering
\begin{tabular}{|c|c||c|c|c|}
\hline
\multicolumn{2}{|c||}{\textbf{Complete Dataset}} & \multicolumn{1}{c|}{\textbf{MCAR}} & \multicolumn{1}{c|}{\textbf{MAR}}& \multicolumn{1}{c|}{\textbf{MNAR}}\\
\hline
\textbf{IQ} & \textbf{Ratings} & \textbf{Ratings} & \textbf{Ratings} & \textbf{Ratings} \\
\hline\hline
78 & 9 & ? & ? & 9 \\\hline
84 & 13 & 13 & ? & 13 \\\hline
84 & 10 & ? & ? & 10 \\\hline
85 & 8 & 8 & ? & ? \\\hline
87 & 7 & 7 & ? & ? \\\hline
91 & 7 & 7 & ? & ? \\\hline
92 & 9 & 9 & 9 & 9 \\\hline
94 & 9 & 9 & 9 & 9 \\\hline
94 & 11 & 11 & 11 & 11 \\\hline
96 & 7 & ? & 7 & ? \\\hline
99 & 7 & 7 & 7 & ? \\\hline
105 & 10 & 10 & 10 & 10 \\\hline
105 & 11 & ? & 11 & 11 \\\hline
106 & 15 & 15 & 15 & 15 \\\hline
108 & 10 & 10 & 10 & 10 \\\hline
112 & 8 & ? & 8 & ? \\\hline
113 & 12 & 12 & 12 & 12 \\\hline
115 & 14 & 14 & 14 & 14 \\\hline
118 & 16 & 16 & 16 & 16 \\\hline
134 & 12 & ? & 12 & 12 \\\hline

\end{tabular}
    \caption{Example from Enders~\cite{Missing_Mechanisms}}
\label{tab:missing_mech}
\end{table}

The missing mechanism refers to the underlying process that generates missing values in a dataset. The missing mechanism essentially describes the rule or relationship between $X^o$ and $X^m$. Understanding the missing mechanism is crucial because it reveals the potential complexities and provides insights into the hidden information that can help us understand the statistical relationships within the data. Table \ref{tab:missing_mech} shows a missing mechanisms example proposed by~\cite{Missing_Mechanisms}. This table \ref{tab:missing_mech} contains two variables: IQ and Job Performance Ratings. The complete data sorted by IQ is in the left two columns. Missing values for Job Performance Ratings under different mechanisms are shown in the right three columns. The symbol '?' represents a missing value in each cell. For MCAR (Missing Completely At Random) data, random rating values are missed (i.e., there is no specific mechanism governing missingness). In the MAR (Missing At Random) data, all cases with missing job performance ratings belong to participants with a lower \textit{IQ} (i.e., IQ value determines the missingness of Ratings). For MNAR (Missing Not At Random) data, all Rating values lesser than 9 are missing (i.e., some specific rating values are missing, they are not dependent on IQ values but some condition on rating itself).

\subsubsection{Missing Completely at Random (MCAR)}
Missing data in the MCAR mechanism implies that the missingness is unrelated to the specific observation under study or any other variables within the dataset. In other words, there are no systematic differences between the observations with missing data and those without missing data.
\\\\
Mathematically, we can express this as:
$$
f(\boldsymbol{M} | \Psi) \textbf{  }  \forall \textbf{  } \boldsymbol{X},\Psi
$$
This expression indicates that the missingness is solely dependent on the parameter $\Psi$ and has no relationship with the observed or missing data itself
\\\\
To illustrate this concept, let's consider an example of MCAR. Suppose we have a table that records various features of employees in a company, such as income, name, gender, and position. Due to a storage shortage, some information is randomly missing across all the features. In this case, the missing values are not related to the salary, gender, or any other information. The missingness is purely random and does not exhibit any systematic patterns or associations with the observed data or any other variables within the dataset.

\subsubsection{Missing At Random (MAR)}
\label{sec:MAR}
In the MAR mechanism, the missingness of data can be predicted based on other variables within the study, but not directly from the missing data itself. In other words, the probability of missingness depends on the observed data, but not on the specific missing values.
\\\\
Mathematically, we can express this as:
$$
f(\boldsymbol{M} | X^o, \Psi) \textbf{  }  \forall \textbf{  } X^m,\Psi
$$
This expression indicates that the missingness of data, represented by $\boldsymbol{M}$, is conditionally dependent on the observed part $X^o$ and the parameter $\Psi$, irrespective of the specific missing values $X^m$.\\\\Returning to the example of employee records, we assume that the missing values occur in the salary feature. Upon further analysis, it is observed that all the missing salary values correspond to female employees. This suggests that the missingness in the salary variable, $X^m$, is related to the observed gender variable, $X^o$. In this scenario, the missingness mechanism is classified as MAR, as the probability of missingness in the salary feature can be predicted based on the observed gender information.

\subsubsection{Missing Not At Random(MNAR)}\label{sec:MNAR}
In contrast to MCAR (Missing Completely At Random) and MAR (Missing At Random), data that are MNAR (Missing Not At Random) exhibit missingness that is directly linked to the value of the missing observation itself, denoted as $\boldsymbol{X}^m$, even when conditioned on the observed part $\boldsymbol{X}^o$. This intrinsic connection suggests that the reasons for data being missing are related to the values that are absent, adding complexity to data analysis.
\\\\For instance, consider a medical study on a sensitive health condition, where patients may be reluctant to disclose information that is stigmatized or deeply personal. If patients with more severe symptoms are less likely to report their status, the missing data (severity of symptoms) is directly related to the unreported value itself. This scenario exemplifies MNAR: the missingness (non-reporting) is not random but tied to the severity of the condition, a critical piece of information for the study.\\\\Handling MNAR data is challenging because standard approaches like deletion or simple imputation can lead to biased results. In the given example, ignoring or incorrectly imputing the missing data could lead to underestimating the severity of the condition, as those with more serious symptoms are underrepresented. Addressing MNAR requires sophisticated techniques, often involving modeling the missing data mechanism itself, to ensure the integrity and accuracy of statistical analyses and research findings. 
\\\\We have outlined the general definition of MNAR missing mechanism. However, it is important to note that the specific missing mechanism under MNAR can vary from case to case, each characterised by different patterns of missingness. For better categorisation, recent work~\cite{Subtypesofthemissing} proposed two distinct subtypes to further divide MNAR into more fine-grained classes.   In this subsection, we delve into these subtypes and provide examples to illustrate their underlying mechanisms.

\subsubsection*{Focused MNAR}
Focused MNAR refers to a scenario where the missingness process depends solely on the missing values $X^m$ and not on the observed values $X^o$. Mathematically, this can be expressed as:$$f(\boldsymbol{M}| X_i^o,\Psi) \textbf{  }  \forall \textbf{  } X_i^m,\Psi$$
Expanding on the previous example, let's consider a situation where an employee with a high salary prefers not to disclose their income and intentionally hides their salary information, resulting in missing values. In this case, the missingness is directly related to the value of $X^m$ itself. 
\subsubsection*{Diffuse MNAR}
Diffuse MNAR occurs when the missingness process involves both missing values and observed values. Mathematically, this can be expressed as:$$f(\boldsymbol{M}| X_i^m,X_i^o,\Psi) \textbf{  }  \forall \textbf{  }X_i^m,\Psi$$
In the case of diffuse MNAR, there is no statistical methodology that can identify the missingness pattern from the observed data alone. Researchers must rely on their judgment and domain knowledge to understand and account for diffuse MNAR.\\\\
To provide an example, consider a survey that includes questions about income, education level, and age. Participants with higher incomes may be less likely to answer questions about income if they also belong to an older generation. On the other hand, younger participants may not exhibit the same reluctance to answer income-related questions, regardless of their income level. Here, the missingness process involves both the observed age values and the unobserved income values, even when conditioned on age.\\\\Additionally, in fields such as meteorology, missing data may arise from outdated wind sensors or equipment failures. For example, wind sensors that are not equipped to record exceptionally high wind speeds may result in missing data during intense hurricanes. However, even newer wind sensors can break under extreme wind conditions. In this case, the missingness process depends on both the unobserved wind speeds and the observed age of the equipment.

\section{Related Work} 
\label{sec:Related Work}
In the field of data analysis, incomplete data sets are a pervasive problem with various underlying causes. These missing data can arise due to constraints in data collection, malfunctioning data acquisition equipment, or non-responsive survey participants~\cite{1}. Little \& Rubin~\cite{little} have extensively studied the prevalence of incomplete data, emphasizing the need for robust statistical approaches to effectively address this issue.
\\\\Several review articles, surveys, and books have been published on the topic of missing data. For instance, articles like \cite{healthcareDL} have discussed imputation approaches in the healthcare field. Norazian et al. \cite{medical2013roles} focus specifically on imputation methods and software for handling missing data in time-series datasets. Garcia et al. \cite{garcia2010pattern} analyze the missing data problem in pattern classification tasks, comparing different methods including ensemble and fuzzy approaches. Other studies, such as Velasco et al. \cite{velasco2020real}, delve into missing values in sensor data of marine systems. Emmanuel et al. \cite{emmanuel2021survey} and Jegadeeswari et al. \cite{jegadeeswari2022missing} discuss machine learning approaches for handling missing data, but do not specifically explore deep learning methods. Adhikari et al. \cite{adhikari2022comprehensive} focus on missing data in IoT, though special missing mechanisms are not explicitly addressed. Liu et al.\cite{liu2023handling} and Sun et al provides missing values handling review in healthcare. Ma et al. \cite{ma2018bayesian} proposed a Bayesian-based imputation method survey paper, while Graham et al. \cite{graham2009missing} studied the utilization of normal-model multiple imputation. Dong et al. \cite{dong2013principled} compared multiple imputation, full information maximum likelihood, and expectation-maximization algorithms. Sun et al. \cite{dlreview} conducted a review and comparative study on deep learning and traditional machine learning methods.
Pereira et al. \cite{pereira2020reviewing} presents a review of using autoencoders for missing data imputation. There are also other review that focus on other areas \cite{pigott2001review,  baraldi2010introduction, lin2020missing, 10225306}
\\\\Despite the existing literature, there is currently no survey study specifically examining the utility of imputation methods for handling missing data with different special missing mechanisms. This survey aims to bridge this gap by providing a comprehensive analysis of imputation approaches, computing platforms, and new perspectives for handling missing data. We go beyond the scope of previous surveys to address the challenges posed by special missing mechanisms and compare their utility in practical applications.

\subsection{Contribution}

In this review, our main contributions can be summarized as follows:

\begin{enumerate}

\item We summarize and discuss various methods for handling missing data with special missing mechanisms. As datasets become increasingly large and complex, traditional statistical-based and machine learning-based methods may not be powerful enough to address these challenges. By focusing on deep learning-based approaches, we aim to provide researchers and practitioners with a valuable resource for effectively handling missing data in diverse domains.

\item We summarize the different missing data generating methods from existing literature. As mentioned, most approaches only utilize MCAR missing mechanism, only small proportion of method mentioned about MAR and MNAR. For those methods consider MAR and MNAR, they usually use different generation method which made us hard to compare. 

\item We suggest future research directions to overcome the limitations of existing imputation techniques and enhance the adoption of advanced methods in practical settings. By identifying gaps in the literature and highlighting new applications for imputation schemes, we aim to guide researchers and practitioners in implementing imputation approaches for different types of data and advancing the field of missing data handling.
\end{enumerate}

Overall, this review serves as a comprehensive resource for researchers and practitioners in the field of missing data handling. By consolidating the knowledge on generating missing data with special missing mechanisms and summarizing deep learning-based imputation methods, we aim to facilitate the development of more effective and reliable techniques for handling missing data in various domains.

\section{Existing Methods of Handling Missing Data}
\label{sec:Existing Methods of Handling Missing Data}
\subsection{Taxonomy}
In this work, we will work on some typical methods that can handle all types of missing mechanisms and also expand their ability to the special missing mechanisms. Table \ref{fig:taxtable}  and  Figure \ref{fig:commonmethods} illustrate the primary taxonomy of methods used to handle missing data and their citations. These methods can be broadly categorized into three groups: \textbf{Deletion} methods, \textbf{Imputation} methods, and \textbf{Representation learning} method.

\begin{itemize}
    \item \textbf{Deletion}\\Deletion methods are the most straightforward approach for handling missing data, where rows or columns containing missing values are simply removed from the dataset. While deletion methods are easy to understand and implement, they can lead to biased outcomes, especially when dealing with missing data that follows special missing mechanisms. The indiscriminate removal of data can result in a loss of valuable information and may introduce biases in subsequent analyses.
    \item \textbf{Imputation} \\ To address the limitations of deletion methods, imputation methods play a vital role in handling missing data. Imputation techniques aim to recover the missing values while preserving the integrity of the complete dataset. These methods are precious when the available data samples are limited or when dealing with missing data characterized by special missing mechanisms. Imputation methods involve filling in the missing values using various strategies, such as mean imputation, regression imputation, and machine learning techniques. By leveraging the information from the observed data, imputation ensures a more comprehensive analysis and reduces the potential biases introduced by data removal.\\
In this review, our primary focus is on imputation methods. Figure \ref{fig:Partial Imputation Method by Year (2005-2023)} provides keyword search results from the Scopus database of different imputation methods (spanning the timeframe from 2000 to 2023). The critical aspect of our investigation lies in the realm of neural network-based approaches for imputation. Figure \ref{fig:Neural Network Based Imputation Method by Year} shows the trends for network-based imputations among all data types. 
    \item \textbf{Representation learning}\\
Representation learning, or feature learning, is a powerful approach in machine learning that automatically learns meaningful representations or features from raw data. These learned representations enable more effective downstream tasks by capturing the underlying structure and patterns in the data. Representation learning methods can be employed to impute missing data by leveraging the learned representations in handling missing values. Instead of relying solely on raw data for imputation, the learned representations are utilized to improve the quality and accuracy of imputed values. The advantage of representation learning lies in its ability to uncover complex relationships and dependencies in the data, which can enhance the imputation process and lead to more robust and reliable results. Consequently, representation learning offers a promising avenue for addressing missing data challenges. Sometimes, representation learning is combined with other imputation techniques or used as a comprehensive imputation strategy.
\end{itemize}

\begin{table}
\centering
\begin{tabular}{|c||c||c||c|}

\hline
     Method&Subtype&Subtype & Citation  \\\hline\hline
     \multirow{2}*{Delete}& Pair-wise& - &~\cite{1,little}\\
     \cline{2-4}
       ~& List-wise& - & ~\cite{1,little}\\\hline\hline\hline

     \multirow{5}*{Imputation}& \multirow{2}*{Statistical-based SI}& Mean/Median/Mode &~\cite{regression1,houari2014handling,hadeed2020imputation} \\\cline{3-4}
         ~& & LOCF \& NOCB &~\cite{LOCF,1,kenward2009last}  \\\cline{2-4}   
    -& \multirow{3}*{Statistical-based MI}& Maximum Likelihood &~\cite{OriginalEM,EM} \\\cline{3-4}
    &  & Matrix Completion &\cite{pca1,qu2009ppca,johnson1990matrix,hernandez2014probabilistic}  \\\cline{3-4}
    &  & Bayesian Approach &~\cite{galimard2018heckman,lin2010comparison,takahashi2017statistical,MICE} \\\hline\hline

    \multirow{5}*{Imputation}& \multirow{5}*{Machine Learning}& Regression-based &~\cite{regression1,regression2,regression3}  \\\cline{3-4}
    & & KNN-based &~\cite{knn1,knn2,knn3,knn4} \\\cline{3-4}
    & & Tree-based &\cite{rf,deza2006dictionary,dt1, dt103}   \\\cline{3-4}
    & & SVM-based &~\cite{svmreg,svm79,svm80,svm84}\\\cline{3-4}
    & & Clustering-based &~\cite{clustering98,clustering102,clustering,zhang2006clustering}  \\\hline\hline

    \multirow{5}*{Imputation}& \multirow{5}*{Neural Network}
    & ANN-based& ~\cite{ann62,ann66,annmlp}\\\cline{3-4}
& & Flow-based& ~\cite{EMflow,mcflow} \\\cline{3-4}
    & & VAE-based &~\cite{miwae,notmiwae,GINA,GPVAE,MCVAE,partialVAE}  \\\cline{3-4}
    & & GAN-based & ~\cite{gan2,gain,misgan,gamin,GI,GANsurvey} \\\cline{3-4}
    & & Diffusion-based&~\cite{tabcsdi,csdi}\\\hline\hline

    \multirow{3}*{Imputation}& \multirow{3}*{Optimization Algorithms}
    & Data Enhanced &~\cite{GA74,GA80,GA81,GA82}  \\\cline{3-4}
    &  & Training Enhanced & ~\cite{optimaltransport} \\\cline{3-4}
    &  & Hyperparemeter Enhanced&~\cite{jarrett2022hyperimpute}\\\hline\hline\hline
    
    Representation & Graph Neural Networks& -&~\cite{representationlearning_GNN,representationlearning_GNN_ehr,representationlearning_gnn_spatio} \\\cline{2-4}
   Learning & AutoEncoder& -&~\cite{representationlearning_AE_imputation,representationlearning_autoencoder,representationlearning_autoencoder_video} \\\hline

\end{tabular}

\caption{Overview of Different Method of Handling Missing Data and Their Citations. MI and SI means Multiple Imputation and Single Imputation}

    \label{fig:taxtable}
\end{table}

\usetikzlibrary{trees,positioning,shapes,shadows,arrows.meta}

\definecolor{g1}{HTML}{eaf2ed}
\definecolor{g2}{HTML}{c5ddcf}
\definecolor{g3}{HTML}{9fc9b1}
\definecolor{g4}{HTML}{79b493}
\definecolor{g5}{HTML}{539f75}
\definecolor{g6}{HTML}{2e8b57}

\definecolor{b1}{HTML}{f0f0f1}
\definecolor{b2}{HTML}{dfe4ec}
\definecolor{b3}{HTML}{cdd7e6}
\definecolor{b4}{HTML}{bccbe1}
\definecolor{b5}{HTML}{aabedc}
\definecolor{b6}{HTML}{99b2d6}
\definecolor{b7}{HTML}{88a5d1}
\definecolor{b8}{HTML}{7799cc}

\definecolor{bb1}{HTML}{c8d5de}
\definecolor{bb2}{HTML}{a1baca}
\definecolor{bb3}{HTML}{7a9fb6}

\definecolor{r1}{HTML}{f1f0f0}
\definecolor{r2}{HTML}{e6d2ce}
\definecolor{r3}{HTML}{dab5ad}
\definecolor{r4}{HTML}{ce978c}
\definecolor{r5}{HTML}{c37a6b}
\definecolor{r6}{HTML}{b75c49}

\begin{figure}
    \centering

\tikzset{
    basic/.style  = {draw, text = black, text width=2cm, align=center, font=\sffamily, rectangle},
    root/.style   = {basic, rounded corners=2pt, thin, align=center, fill=b1},
    deletion1/.style = {basic, thin, rounded corners=2pt, align=center, fill=g1,text width=2cm,},
    deletion2/.style = {basic, thin, align=left, fill=g3, text width=1.5cm},
    imputation1/.style = {basic, thin, rounded corners=2pt, align=center, fill=b2,text width=2cm,},
    imputation2/.style = {basic, thin, rounded corners=2pt, align=center, fill=b3,text width=2.5cm,},
    imputation_Stats/.style = {basic, thin, rounded corners=2pt, align=center, fill=bb1,text width=1.5cm,},
    imputation_Stats_c/.style = {basic, thin, rounded corners=2pt, align=center, fill=bb2,text width=2.5cm,},
    imp_main/.style = {basic, thin, rounded corners=2pt, align=center, fill=b4,text width=2.3cm,},
    imp_main_d/.style = {basic, thin, rounded corners=2pt, align=center, fill=b6,text width=2.3cm,},
    opleaf/.style = {basic, thin, rounded corners=2pt, align=center, fill=b7,text width=3cm,},
    edge from parent/.style={draw=black, edge from parent fork right},
    rep1/.style = {basic, thin, rounded corners=2pt, align=center, fill=r2,text width=3cm,},
    rep2/.style = {basic, thin, rounded corners=2pt, align=center, fill=r3,text width=2.3cm,},
}

\begin{forest}for tree={
    grow=east,
    growth parent anchor=west,
    parent anchor=east,
    child anchor=west,
    edge path={\noexpand\path[\forestoption{edge},->, >={latex}] 
         (!u.parent anchor) -- +(10pt,0pt) |-  (.child anchor) 
         \forestoption{edge label};}
}
[Handling Missing Data, root,  l sep=10mm,
    [\textbf{Representation Learning}\\Section \ref{sec:Representation Learning for Handling Missing Data}, rep1,  l sep=5mm,
        [Graph Neural Networks, rep2]
        [Auto-Encoder, rep2]
        ]
    [\textbf{Imputation}\\Section \ref{sec:imputation methods}, imputation1,  l sep=5mm,
        [\textbf{Optimization Algorithm}\\Section \ref{sec:Optimization}, imputation2,  l sep=5mm, 
            [Hyperparameters-Enhanced, opleaf ]
            [Training-Enhanced, opleaf ]
            [Data-Enhanced, opleaf ]
    ]
        [\textbf{Neural Network}\\Section \ref{sec:Neural Network}, imputation2,  l sep=10mm,
            [Diffusion, imp_main_d ]
            [GAN, imp_main_d ]
            [VAE, imp_main_d ]
            [Flow, imp_main_d ]
            [ANN, imp_main_d ]    
        ] 
        [\textbf{Machine Learning}\\Section \ref{sec:Machine Learning}, imputation2,  l sep=10mm,
            [Clustering, imp_main ]
            [SVM, imp_main ]
            [Tree, imp_main ]
            [KNN, imp_main ]
            [Regression, imp_main ]
        ]
        [\textbf{Statistical}\\Section \ref{sec:Statistical}, imputation2,  l sep=10mm,
            [\textbf{Multiple}\\Section \ref{sec:MI}, imputation_Stats,  l sep=5mm,
                [Bayesian, imputation_Stats_c ]
                [Matrix Completion, imputation_Stats_c ]
                [Maximum Likelihood, imputation_Stats_c ]
        ] 
            [\textbf{Single}\\Section \ref{sec:SI}, imputation_Stats,  l sep=5mm,
                [LOCF \& NOCB,imputation_Stats_c]
                [Mean\\Median\\Mode,imputation_Stats_c]
                [0 Imputation,imputation_Stats_c]
            ]
        ]
    ]
    [\textbf{Deletion}\\ Section \ref{sec:Deletion for Handling Missing Data}, deletion1,  l sep=5mm,
        [Pair-wise, deletion2]
        [List-wise, deletion2]
    ]      
]
\end{forest}
    \caption{Method Taxonomy for Handling Missing Data}
    \label{fig:commonmethods}
\end{figure}
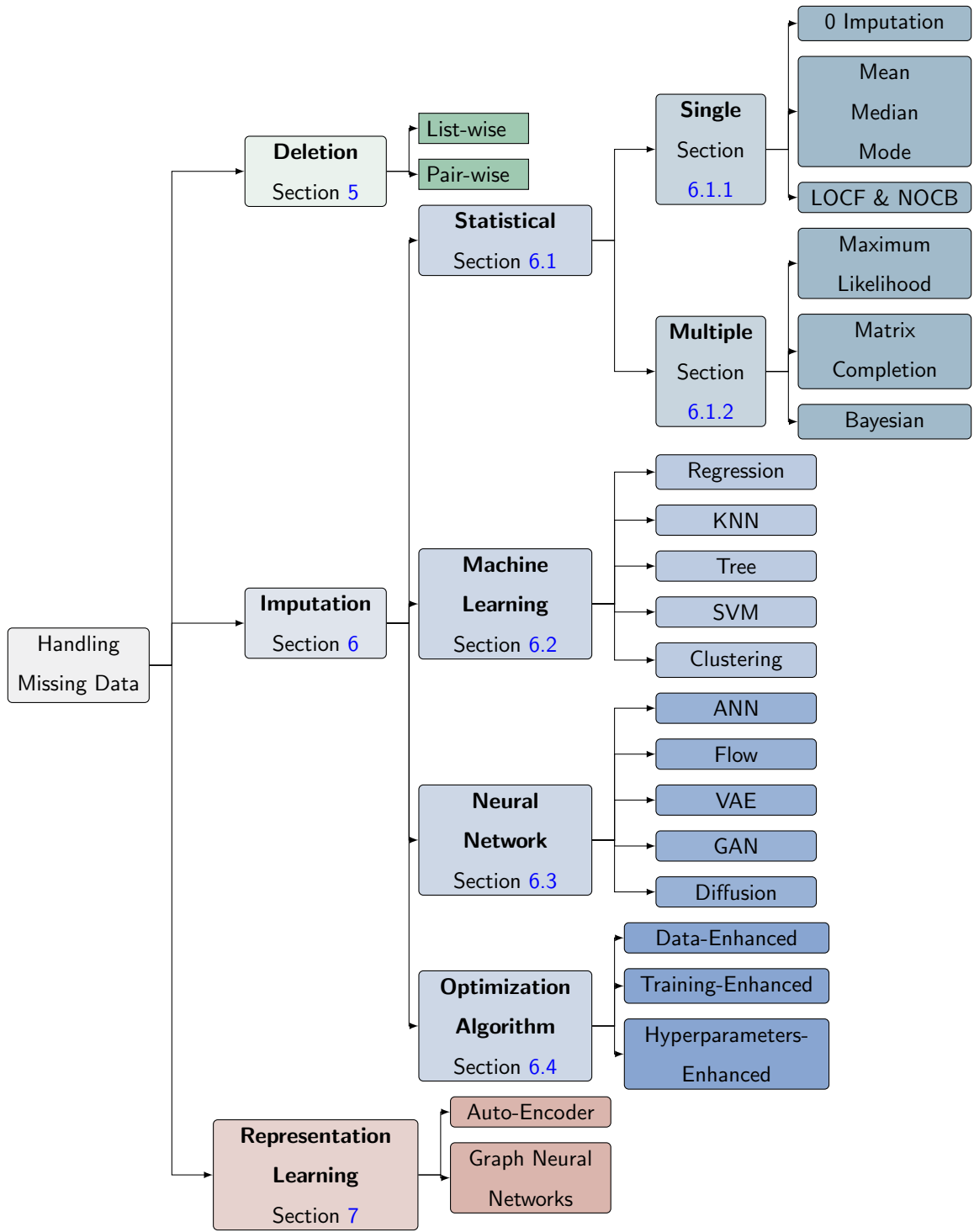

\usetikzlibrary{trees,positioning,shapes,shadows,arrows.meta}

\pgfplotsset{compat=1.16} 

\pgfplotsset{
    bar group size/.style 2 args={
        /pgf/bar shift={%
                -0.5*(#2*\pgfplotbarwidth + (#2-1)*\pgfkeysvalueof{/pgfplots/bar group skip})  + 
                (.5+#1)*\pgfplotbarwidth + #1*\pgfkeysvalueof{/pgfplots/bar group skip}},%
    },
    bar group skip/.initial=2pt,
    plot 0/.style={c2,fill=c1,mark=none},%
    plot 1/.style={c4,fill=c3,mark=none},%
    plot 2/.style={c6,fill=c5,mark=none},%
    plot 3/.style={c8,fill=c7,mark=none},%
}

\definecolor{c1}{HTML}{a4cc90}
\definecolor{c2}{HTML}{7cb990}
\definecolor{c3}{HTML}{58a590}
\definecolor{c4}{HTML}{3f908e}
\definecolor{c5}{HTML}{277a8b}
\definecolor{c6}{HTML}{1c6388}
\definecolor{c7}{HTML}{254a7f}
\definecolor{c8}{HTML}{2c3071}

\definecolor{b1}{HTML}{f0f0f1}
\definecolor{b2}{HTML}{dfe4ec}
\definecolor{b3}{HTML}{cdd7e6}
\definecolor{b4}{HTML}{bccbe1}
\definecolor{b5}{HTML}{aabedc}
\definecolor{b6}{HTML}{99b2d6}
\definecolor{b7}{HTML}{88a5d1}
\definecolor{b8}{HTML}{7799cc}
\definecolor{bb1}{HTML}{c8d5de}
\definecolor{bb2}{HTML}{a1baca}
\definecolor{bb3}{HTML}{7a9fb6}

\begin{figure}[!h]
    \centering

\begin{tikzpicture}[yscale=0.7, xscale = 0.7]
    \begin{axis}[
        width=1*\textwidth,
        x tick label style={/pgf/number format/1000 sep=},
        enlargelimits=0.05,
        ymax=50, 
        legend pos=north west, 
        ybar,
        bar width=4pt,
        xtick={2000,2001,2002,2003,2004,2005,2006,2007,2008,2009,2010,2011,2012,2013,2014,2015,2016,2017,2018,2019,2020,2021,2022,2023},
        xticklabel style={rotate=45, anchor=north east}
        ]

        \addplot[plot 1,bar group size={0}{3}]
            coordinates {           
           (2005,1)(2003,1) (2007,1)(2009,1) (2010,3)(2013,1) (2014,1)(2015,1)(2016,1) (2020,1) (2018,1)(2021,1)(2022,3)(2023,1)
            };
        \addplot[plot 0,bar group size={1}{3}]
         coordinates {
(2023,	15)
(2022,	28)
(2021,	14)
(2020,	16)
(2019,	11)
(2018,	16)
(2017,	18)
(2016,	14)
(2015,	21)
(2014,	16)
(2013,	12)
(2012,	8)
(2011,	9)
(2010,	10)
(2009,	6)
(2008,	7)
(2007,	5)
(2006,	5)
(2005,	2)
(2004,	4)
(2003,	1)
(2002,	1)
(2000,	1)
         };

        \addplot[plot 2,bar group size={2}{3}]
        coordinates {
(2023,	29)
(2022,	41)
(2021,	30)
(2020,	27)
(2019,	16)
(2018,	17)
(2017,	13)
(2016,	5)
(2015,	10)
(2014,	1)
(2013,	10)
(2012,	10)
(2011,	4)
(2010,	4)
(2009,	4)
(2008,	3)
(2006,	2)
(2005,	1)
(2004,	2)
(2003,	1)

};
        \addplot[plot 3,bar group size={3}{3}]
        coordinates {
(2023,	22)
(2022, 18)
(2021, 15)
(2020, 10)
(2019, 8)
(2018,	1)
((2015,	2)
(2014	,2)
(2012,	1)
};

        \legend{Statistical-Single,Statistical-Multiple,Machine Learning,Neural Network}
    \end{axis}
\end{tikzpicture}
    \caption{Keyword search from Scopus database for MNAR and MAR Data Imputation (All data  types) Method by Year(2000-2023)}
    \label{fig:Partial Imputation Method by Year (2005-2023)}

\end{figure}
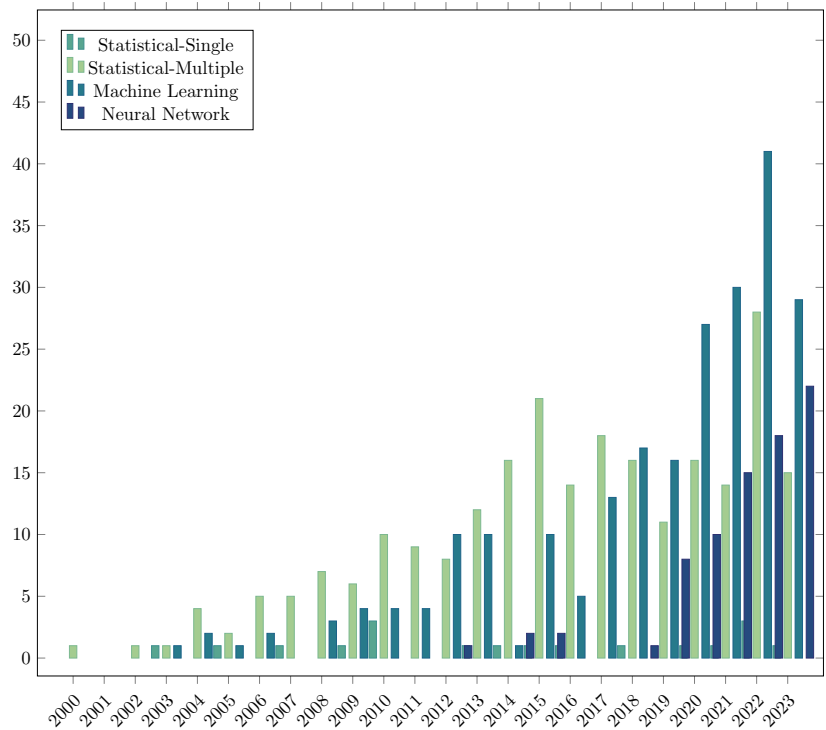


\pgfplotsset{compat=1.16} 

\definecolor{c1}{HTML}{a4cc90}
\definecolor{c2}{HTML}{7cb990}
\definecolor{c3}{HTML}{58a590}
\definecolor{c4}{HTML}{3f908e}
\definecolor{c5}{HTML}{277a8b}
\definecolor{c6}{HTML}{1c6388}
\definecolor{c7}{HTML}{254a7f}
\definecolor{c8}{HTML}{2c3071}

\definecolor{b1}{HTML}{f0f0f1}
\definecolor{b2}{HTML}{dfe4ec}
\definecolor{b3}{HTML}{cdd7e6}
\definecolor{b4}{HTML}{bccbe1}
\definecolor{b5}{HTML}{aabedc}
\definecolor{b6}{HTML}{99b2d6}
\definecolor{b7}{HTML}{88a5d1}
\definecolor{b8}{HTML}{7799cc}
\definecolor{bb1}{HTML}{c8d5de}
\definecolor{bb2}{HTML}{a1baca}
\definecolor{bb3}{HTML}{7a9fb6}

\pgfplotsset{
    bar group size/.style 2 args={
        /pgf/bar shift={%
                -0.5*(#2*\pgfplotbarwidth + (#2-1)*\pgfkeysvalueof{/pgfplots/bar group skip})  + 
                (.5+#1)*\pgfplotbarwidth + #1*\pgfkeysvalueof{/pgfplots/bar group skip}},%
    },
    bar group skip/.initial=2pt,
    plot 0/.style={c2,fill=c1,mark=none},%
    plot 1/.style={c4,fill=c3,mark=none},%
    plot 2/.style={c6,fill=c5,mark=none},%
    plot 3/.style={c8,fill=c7,mark=none},%
}

\begin{figure}[!h]
    \centering

\begin{tikzpicture}[yscale=0.7, xscale = 0.7]
    \begin{axis}[
        width=0.9*\textwidth,
        x tick label style={/pgf/number format/1000 sep=},
        enlargelimits=0.15,
        ymax=30, 
        ybar,
        bar width=10pt,
        xtick={2017,2018,2019,2020,2021,2022,2023},
        ytick={0, 10, 20, 30}
        ]
        

        \addplot[plot 0,bar group size={0}{3}]
         coordinates {(2017,1) (2018,2) (2019,3) (2020,9) (2021,13)  (2022,25) (2023,17) };
        \addplot[plot 1,bar group size={1}{3}]
        coordinates {(2018,1) (2019,3) (2020,9) (2021,8) (2022,12) (2023,6) };
        \addplot[plot 2,bar group size={2}{3}]
        coordinates { (2019,1) (2020,1) (2021,1)};
        \addplot[plot 3,bar group size={3}{3}]
        coordinates { (2020,1) (2021,4)  (2022,7) (2023,11)};

        \legend{GAN,VAE,Flow,Diffusion}
    \end{axis}
\end{tikzpicture}
    \caption{Keyword search from Scopus database for Neural Network Based Data Imputation (All data  types) Method by Year}
    \label{fig:Neural Network Based Imputation Method by Year}

\end{figure}
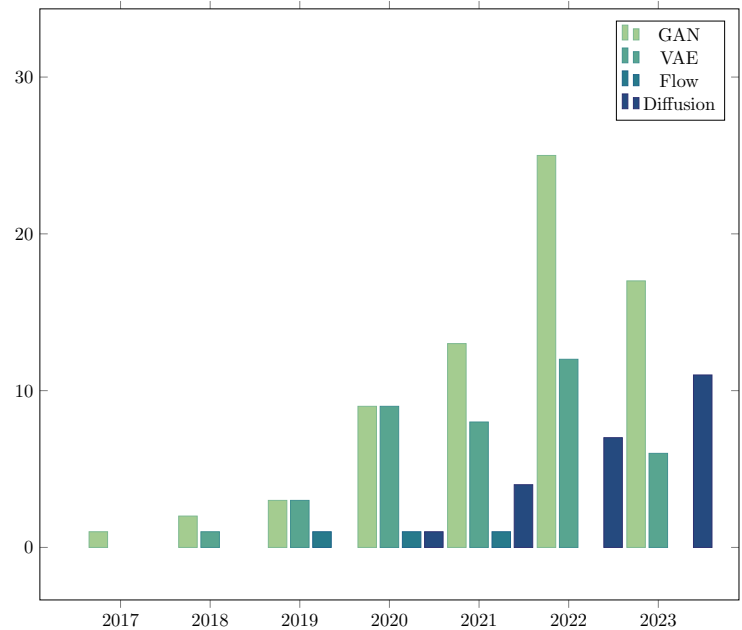


\section{Deletion for Handling Missing Data}
\label{sec:Deletion for Handling Missing Data}
\subsection{Listwise Deletion}
\label{sec:Listwise Deletion}
Listwise deletion, also known as complete-case analysis, is a commonly used method in data analysis. It involves removing any case with one or more missing values from the dataset. For example, if the $1st$ instance has a missing value in the $jth$ column, the entire $1st$ instance will be removed from the analysis.

\subsection{Pairwise Deletion}
\label{sec:Pairwise Deletion}
Pairwise deletion, referred to as available-case analysis, handles missing values by removing cases containing missing values only when the affected features are used in analyses. For example, if the $1st$ instance has a missing value in the $jth$ column, the $1st$ instance will be excluded only when analyzing the $jth$ feature. However, it will be included in analyses involving other features.
\\\\
Both listwise and pairwise deletion methods work well when the missing data follow the MCAR assumption, where the missingness is unrelated to the observed and unobserved variables. In MCAR scenarios, the complete data can be considered a simple random sample from the original target population. However, when the missing data follows special missing mechanisms, such as MNAR, deletion methods can lead to biased parameter estimates and compromise the validity of the analysis.
\\\\
For example, suppose all the missing data occurs in individuals from high-income groups, and those instances are deleted. In that case, the resulting dataset will only represent individuals from low-income groups, introducing bias and potentially misleading results. Additionally, using deletion methods reduces the adequate sample size, which can lead to a decrease in statistical power and precision.
\\\\
It is essential to be cautious when using deletion methods and consider their limitations. These methods discard valuable information in the missing values, potentially resulting in losing important insights. Alternative approaches, such as imputation methods, can be explored to handle missing data more effectively and mitigate the limitations associated with deletion methods.


\section{Imputation for Handling Missing Data}
\label{sec:imputation methods}
Imputation approaches encompass various techniques. In this section, we divide the imputation method into several subsections, including statistical-based methods, machine learning-based methods, deep learning-based methods, and optimization methods. These techniques are used to fill in missing values in the dataset. We will introduce the imputation method for missing data also analysis they ability for handling missing data for special miss mechanism.

\subsection{Statistical Based Imputation}
\label{sec:Statistical}
Statistical imputation is classified into two types as Single Imputation(SI) and Multiple Imputation(MI). 

\subsubsection{Single Imputation}
\label{sec:SI}
Single imputation involves the replacement of missing values with a single estimated value. In this approach, each missing value is imputed with a single value, utilizing specific assumptions or statistical methods. This process ensures that each missing value is assigned a single imputed value, resulting in a one-time imputation.

\begin{itemize}
    \item \textbf{Mean, Median and Mode Imputation}\\
Mean, median, and mode imputation methods involve calculating the mean, median, or mode of non-missing values within each column and using these values to impute missing values~\cite{regression1}. Mean and median imputation are suitable for numerical data, while mode imputation is more appropriate for categorical or binary data, as they do not have mean or median values. These methods can be applied not only to tabular datasets but also to other formats such as image data and time series data, as they can be represented in numerical formats. However, it is important to note that since these methods rely on simple calculations, they may not capture the complex underlying structure of the data distribution.
\\\\While mean, median, and mode imputation methods are easy to implement, they have certain limitations. They are effective only if the missing mechanism follows the MCAR assumption~\cite{regression1}. Mean imputation can lead to overestimation of the sample size, underestimation of the variance, and negatively biased correlations~\cite{houari2014handling}. Despite these drawbacks, in certain datasets and scenarios, mean imputation may still outperform other imputation techniques\cite{hadeed2020imputation}.\\\\However, it is important to note that mean, median, and mode imputation methods may not be suitable for handling missing data with special missing mechanisms, as they do not consider the unique patterns and complexities associated with such mechanisms~\cite{houari2014handling}. These methods provide a simple solution but may not capture the full complexity of the missing data problem. Therefore, alternative imputation methods specifically designed for handling special missing mechanisms should be considered in such cases.
\item \textbf{LOCF and NOCB}

Last Observation Carried Forward (LOCF) and Next Observation Carried Backward (NOCB) methods involve using the nearest observed value before or after the missing value to impute the missing value ~\cite{LOCF}. These methods are commonly used in longitudinal and time-series datasets and also related to missing pattern issue, where variables are repeatedly measured over a series of time points or observation are dependent with the nearest points. LOCF imputes the missing value with the value of the last observed measurement, while NOCB imputes the missing value with the value of the next observed measurement.\\\\It is important to note that these methods are suitable when the instances or observations have a temporal or sequential relationship. However, if the instances or observations are independent of each other, such as in cross-sectional datasets, also for dealing with special missing mechanisms, these methods may not be appropriate.
\end{itemize}

\subsubsection{Multiple Imputation}
\label{sec:MI}
Multiple Imputation is a widely recognized approach for handling missing data, initially proposed by Rubin~\cite{rubin2004multiple}. It involves replacing a missing value with more than one possible value, typically using statistical models and algorithms. Multiple imputation generates multiple completed datasets, where each dataset contains imputed values for the missing data. These imputed datasets are then analyzed using standard statistical methods, and the results are combined to obtain valid inferences and estimates.
\\\\
Multiple imputation offers several advantages over single imputation methods. By generating multiple imputed datasets, it takes into account the uncertainty associated with imputing missing values. It also captures the variability between imputations, resulting in more accurate estimates and appropriate measures of uncertainty. There are various methods and algorithms available for implementing multiple imputation. In this section, some of the well-known approaches include Maximum Likelihood methods, Matrix Completion Methods and Bayesian Methods will be listed.
These methods differ in their assumptions, modeling techniques, and handling of specific data types. The choice of method depends on the characteristics of the dataset, the missing data mechanism, and the research objectives.

\begin{itemize}
    \item \subsubsection*{Maximum Likelihood Method}

The Maximum Likelihood Imputation (MLI)  is a model-based approach that estimates missing values by maximizing the likelihood function under a specific probabilistic model. The method treats missing values as latent variables and finds the parameter values that make the observed data most probable, given the assumed model. This involves formulating the likelihood function as a product of conditional probabilities, where each missing value's conditional probability is maximized with respect to the unknown parameters. MLI-based method provides consistent imputed values that align with the assumed statistical model and observed data. However, it relies on specific data distribution assumptions, and if the model is not well-suited for the data, the imputed values may be biased. Additionally, MLM can be sensitive to outliers and might not perform optimally with complex missing patterns or high-dimensional data.\\\\
\begin{itemize}
    \item  \textbf{Expectation Maximization (EM) Algorithm}\\
One technique within MLI is the Expectation Maximization (EM) algorithm, introduced by Dempster~\cite{OriginalEM}.  EM algorithm is an iterative algorithm that aims to find maximum likelihood estimates and fit models to missing data problems. Instead of directly imputing missing values, EM estimates the data based on two steps- An expectation step (E-Step) and a maximization step. This process is repeated until the MLI estimates are obtained. The sequence of parameters converges to the MLI estimates, which implicitly average over the distribution of missing values. EM can be used to estimate means, standard deviation, and correlations of interest. However, EM has some limitations. It requires a large sample size and assumes that the missing mechanism is MAR~\cite{EM}. While its convergence is guaranteed under MAR, the convergence rate depends on the fraction of missing data. Low missingness leads to fast convergence, while high missingness leads to slow convergence. Additionally, EM is a complex method, and its convergence can be slow and may yield a sub-optimal results. Thus EM is a good algorithms that can solve MAR data but not MNAR data.
\end{itemize}

\item  \subsubsection*{Matrix Completion Method}
 
Matrix completion is a method used to fill in missing entries in an incomplete matrix that organizes the data. This method is commonly used in recommendation systems.  However, this approach may face limitations when handling diverse data formats. For instance, irregular or unstructured data may not be adequately represented using a fixed-size matrix. Similarly, timeseries data with varying lengths can pose challenges in fitting them into a matrix. Additionally, a matrix with a large number of missing values becomes sparse, making it challenging to deal with severe missing data. Despite these limitations, matrix completion remains a valuable data imputation approach. It includes common methods such as Principal Component Analysis (PCA), Probabilistic Principal Component Analysis (PPCA), and Probabilistic Matrix Factorization (PMF). These methods aim to make assumptions about the data matrix to create a well-posed problem, such as maximizing determinant, positive definiteness, or low rank. By assuming a low-rank structure, which implies correlations between entries, missing entries can be recovered using convex optimization when there are enough observed entries.
\begin{itemize}
    \item \textbf{Principal Component Analysis} \\Principal Component Analysis (PCA) is a dimensionality reduction technique used in multivariate statistics and machine learning. Its primary objective is to transform a high-dimensional dataset into a lower-dimensional representation while preserving as much of the original data's variance as possible. In Philip et al. ~\cite{pca1}'s study, the PCA approach adopted is akin to the projection to the model plane introduced. Before constructing the initial PCA model from the available data, the appropriate number of components is estimated to ensure optimal dimension reduction. As for handling missing data, they employ a method inspired by the idea of PCA, which involves the development of a regression model called Partial Least Squares (PLS). This PLS-based regression model is utilized for imputing missing values, effectively enhancing the completeness of the dataset. The imputation process iteratively substitutes missing values with imputed values generated using regression loadings from the updated data matrix, ensuring a robust and accurate completion of the missing data ~\cite{pca1}.\\The missing data imputation method based on PCA is usually rooted in a strong theoretical framework, leading to improved algorithm performance when applied to mixed datasets. However, this approach faces challenges, such as high computational complexity, potential convergence issues, and the risk of overfitting. Instability in the imputations may also arise when dealing with datasets containing numerous missing values. Despite its merits, PCA alone has some limitations for missing data imputation. PCA has been extended through various methods, such as PPCA~\cite{qu2009ppca} and Bayesian Principal Component Analysis (BPCA), to address these limitations and enhance the imputation process. PPCA combines EM and PCA techniques, while BPCA incorporates Bayesian estimation and PCA. 
    \item \textbf{Probabilistic Principal Component Analysis} \\
The Probabilistic Principal Component Analysis-based (PPCA) missing data imputation method incorporates two key techniques, namely PCA and Maximum Likelihood Estimation (MLE). This approach uses PCA to distinguish between significant and dominant components of the traffic flow, effectively separating them from trivial and un-modelable elements. On the other hand, MLE is employed to estimate the missing values based on the identified dominant components~\cite{qu2009ppca}.\\ From a latent variable analysis perspective, PPCA captures the statistical characteristics of the known data, indirectly constructing a latent sliding regression model. This strategy leverages the redundancy in flow data to establish a lower intrinsic dimensionality of the observed time series. The latent model is concurrently built as the missing values are gradually recovered. This process strikes a favourable balance between periodicity, local predictability, and other statistical properties of the traffic flow, making it effective in outperforming conventional methods, particularly when dealing with relatively high missing data ratios.\\By combining the strengths of PCA and MLE, the PPCA-based approach provides a robust and reliable solution for handling missing data. It allows for the efficient imputation of missing values in traffic flow datasets, leading to enhanced data completeness and improved performance compared to traditional methods.
\item \textbf{Probabilistic Matrix Factorization} \\
Probabilistic Matrix Factorization (PMF)~\cite{johnson1990matrix,hernandez2014probabilistic} is a matrix decomposition technique widely used for missing data imputation. It decomposes a data matrix into two lower-dimensional matrices, allowing for the efficient representation of large datasets. PMF offers scalability and robustness against overfitting, making it suitable for various data types, such as continuous, binary, or ordinal data. The method leverages stochastic inference methods to handle large datasets by randomly sub-sampling missing matrix entries, which enhances computational efficiency. However, a key drawback of PMF is its assumption that the missing data is MAR, which means PMF can not solve all missing data with special missing mechanism cases.
\end{itemize}

\item \textbf{Bayesian Approach}\\
In Bayesian-based missing data imputation models, the missing values are treated as unknown parameters drawn randomly from an appropriate probability distribution. The Bayesian paradigm provides a natural and flexible framework to model the uncertainty surrounding the missing values by estimating their posterior distribution given the observed data and any available prior knowledge. The process of Bayesian imputation involves specifying a probabilistic model that captures the relationship between the observed data and the missing values. This model can incorporate various sources of information, such as prior beliefs about the missing values or the underlying data-generating process. By combining the observed data with the probabilistic model, Bayesian methods infer the likely values of the missing data, along with their uncertainty, through posterior estimation.
\begin{itemize}
    \item \textbf{Multivariate imputation by chained equation(MICE)}\\
Multiple Imputation by Chained Equation (MICE Imputation~\cite{MICE}  is a highly flexible and potent missing data imputation method. It utilizes a series of conditional densities to define the multivariate normal distribution for imputing missing values. The process starts with random draws from the corresponding posterior predictive distribution, ensuring proper imputation. At each iteration, random draws are made from the posterior distribution of parameters, similar to the Gibbs sampler. This iterative approach stabilizes the imputed results before finalizing a complete dataset. The original implementation of MICE provides a practical solution for handling MNAR outcomes with MAR predictors using a Heckman selection model within the MICE procedure. The MICE also supports direct application or sensitivity analysis to assess the robustness of MAR mechanisms. Additionally, it offers specialized tools for effectively managing MNAR missing data, empowering researchers to apply these techniques to their datasets.
\item \textbf{Markov chain Monte Carlo}\\
Markov Chain Monte Carlo (MCMC)~\cite{lin2010comparison}\cite{takahashi2017statistical} is a powerful algorithm used for missing data imputation, especially in situations where the missing data problem is complex and multivariate. It approximates the joint posterior distribution when evaluating the true expression for this distribution is analytically difficult or impossible due to arbitrary patterns of missing values and various data types (continuous, nominal, binary, ordinal). The MCMC algorithm operates through iterative I-steps and P-steps. At each iteration, the I-step draws imputations from the current iteration's predictive distribution, considering the pattern of missing variables for each case. The P-step updates the parameter values for the predictive distribution by drawing from the completed data posterior. The MCMC algorithm converges with sufficient iterations, providing imputation draws for the missing values that simulate draws from the true joint posterior distribution. MCMC imputation is based on Bayesian computational algorithms, enabling the estimation of mean, variance, covariance matrix, and other essentials required for imputation. It considers data variability, uses all available data, and provides imputed values as starting points for augmenting missing data points. However, drawbacks include the assumption of multivariate normal distribution, computational demands, and high iteration requirements. 
\end{itemize}

\end{itemize}

\subsection{Machine Learning Based Imputation} 
\label{sec:Machine Learning}
Machine learning-based imputation methods utilize unsupervised or supervised learning to estimate missing values in datasets, leveraging available information from non-missing data for precise predictions. The advantage of machine learning lies in its predictive capability, capturing intricate relationships and patterns within the data. Additionally, these methods are flexible, robust to noise and outliers, and can handle diverse data types, adapting to various missing data patterns. Their effectiveness in reducing bias and handling large datasets makes them a powerful and versatile solution to enhance the accuracy and reliability of analyses involving incomplete data. In later sessions, we will introduce common approaches such as Regression, Classification, and Clustering.
\subsubsection{Regression Based Imputation}
\label{sec:Regression}
Regression is a supervised learning method. The Regression-based missing data imputation methods use regression models to estimate missing values in tabular datasets. These methods create a model using the observed data, treating the variable with missing values as the dependent variable and using other complete variables as predictors~\cite{regression1,regression2,regression3}. The regression model is then utilized to predict the missing values based on the values of the predictor variables. Assume that we let  $X_{ik}$  be the missing value for the $k^{th}$ column of $i^{th}$ instance. The linear regression imputation model will build a model that is shown below:
$$
X_{ik} =  \beta_0 +  \beta_1X_{i1} + \beta_2X_{i2} + \beta_3X_{i3} +,...,\beta_KX_{iK}
$$
Standard regression-based imputation techniques include mean/mode imputation, simple linear regression, multiple imputation and non-linear regression. While linear and logistic regression are suitable for datasets with linear relationships between variables and can handle continuous and categorical data, they may not be ideal for datasets with complex or nonlinear patterns. The accuracy of the imputed values relies on the choice of predictor variables and the performance of the regression model, making careful consideration essential for obtaining reliable imputations. However, in some cases, the regression-based model is not efficient since the regression-based model should always refit the model due to different inputs of missing data and observed data parts.

\subsubsection{K-Nearest Neighbor Based Imputation}
\label{sec:knn}
K-nearest neighbour (K-NN) Imputation, a popular supervised learning method, can also impute missing values in datasets. By selecting the nearest neighbours based on a chosen distance function, K-NN imputes the missing value using the value from the closest neighbor~\cite{knn1,knn2}. The flexibility of this method allows for various distance functions and the number of neighbours, influencing the imputation results~\cite{deza2006dictionary}. K-NN outperforms LOCF and NOCB methods in terms of data type and dimensions. In cases where the missing data follows the MAR mechanism and there is no prior knowledge about its distribution, but information can be gathered from the observed data, K-NN imputation becomes a suitable choice. However, K-NN might not be the most appropriate method for MNAR cases where the missingness is unrelated to the observed data.

\subsubsection{Tree Based Imputation}

Tree-based methods, such as Decision Trees~\cite{dt1, dt103} and Random Forests~\cite{rf}, are widely used supervised learning models for classification and regression tasks. These models utilize partition strategies to construct either a single tree or multiple trees, where the dataset is split into distinct leaves based on input features, with the most crucial feature acting as the root node. Tree-based methods employ feature selection measures like Entropy, Information Gain, and Gini Index to determine the best splitting criteria. Tree-based methods handle missing data imputation by naturally incorporating missing values into their splitting rules during tree-building. These methods can create separate branches for observations with missing values and those with complete data, allowing them to make imputations based on available information.
Moreover, Random Forests, an ensemble technique of multiple decision trees, enhance prediction accuracy and robustness by reducing overfitting and handling noisy data. Tree-based methods show good performance in handling MCAR and MAR values, and they can also handle MNAR values to some extent~\cite{rf}. Additionally, missing data and outliers have minimal influence on decision tree algorithms.

\subsubsection{Support Vector  Machines Imputation}
Support Vector Machine (SVM) is a widely adopted machine learning algorithm for handling missing data~\cite{svmreg,svm79,svm80,svm84}. It seeks to identify an optimal separating hyperplane in a labelled training sample, maximizing the distance between the hyperplane and the nearest data points~\cite{svm80}. \cite{svmreg} employed an SVM regressor for missing data imputation. In contrast, Chechik et al.~\cite{svm84} introduced a novel approach using a max-margin learning framework to address missing values. Their innovative method involved formulating an objective function to maximize the margin of each sample within its specific subspace, resulting in efficient imputation. Their approach saved computational time and demonstrated robustness for MAR missing mechanisms. However, SVM-based imputation lacks extension to MNAR missing mechanisms, and its imputation accuracy is considerably influenced by data type and distribution.

\subsubsection{Clustering Based Imputation}
Clustering, an unsupervised learning technique, groups similar items together based on similarity or distance functions. Common clustering methods such as k-means clustering have been explored for missing data handling in various studies. The k-means method involves randomly assigning centroids and iteratively reallocating data points to the closest centroids to form clusters. This process continues until the assignments stabilize, and then the cluster information is used to handle missing values. \cite{clustering98,clustering,clustering102,zhang2006clustering}. However, the clustering results are influenced by the choice of distance function, the number of clusters (k), and the initial centroid locations. Gajawada et al.~\cite{clustering98} initially proposed using k-means for missing data imputation, but their approach may propagate errors from earlier imputations to subsequent ones. On the other hand, Zhang et al.~\cite{clustering} proposed a clustering-based non-parametric kernel-based imputation technique to handle missing values in target features. The approach demonstrated effectiveness in creating inferences for variance and distribution functions after clustering. However, it did not consider missing values in conditional features and class features. It's essential to carefully consider the choice of clustering methods and their parameters when applying clustering for imputation, as the results heavily rely on these choices.
\\\\
Table \ref{tab:Statisticalbased} provides a summary of various statistical-based and Machine Learning-based Imputation Method imputation methods, along with the suitable data types and missing mechanisms for each method. However, as the size and dimension of datasets continue to grow rapidly, traditional statistical-based and Machine Learning-based imputation methods may encounter challenges in handling large-scale and high-dimensional data. Additionally, the increasing complexity and diversity of data formats and types can further limit the applicability of these methods.
\begin{table}[!h]
    \centering
    \begin{tabular}{|c|c|c|c|c|c|}\hline
        Imputation Method & Numerical&Categorical &MCAR &MAR&MNAR \\\hline\hline

        Listwise Deletion & \cmark & \cmark &\cmark & \xmark & \xmark \\\hline
        Pairwise Deletion & \cmark & \cmark &\cmark & \xmark & \xmark  \\\hline
        Mean/Median  & \cmark & \xmark &\cmark & \xmark & \xmark   \\\hline
        Mode  & \xmark & \cmark &\cmark & \xmark & \xmark   \\\hline
       LOCF \& NOCB  & \cmark & \cmark &\cmark & \xmark & \xmark  \\\hline
       Maximum Likelihood & \cmark & \xmark &\cmark & \cmark & \xmark  \\\hline
       Matrix Completion & \cmark & \cmark &\cmark & \cmark & \cmark  \\\hline

       Bayesian Approach& \cmark & \xmark &\cmark & \cmark & \cmark  \\\hline

        Regression & \cmark & \xmark &\cmark & \cmark & \xmark \\\hline
        K-Nearest Neighbour & \cmark & \cmark & \cmark & \cmark & \xmark \\\hline
        Tree Based & \cmark & \cmark & \cmark & \cmark & \cmark  \\\hline
        SVM Based& \cmark & \xmark & \cmark & \cmark & \cmark   \\\hline
       Clustering Based & \cmark & \cmark &  \cmark & \xmark & \xmark \\\hline
       
    \end{tabular}
    \caption{Summary of Statistical based and Machine Learning-based Imputation Method}
    \label{tab:Statisticalbased}
\end{table}

\subsection{Neural Network Based Imputation Method}
\label{sec:Neural Network}
Neural Network-based methods leverage the power of neural networks to learn complex patterns and impute missing values automatically. These techniques are promising for addressing scalability challenges and providing efficient solutions to handle missing data with special mechanisms in large and complex datasets.

\subsubsection{Artificial Neural Network}

Artificial Neural Networks (ANN) are computational systems inspired by biological neural networks. In the realm of handling missing data in ANNs, there are primarily two distinct approaches. The first, more akin to supervised learning, is exemplified in the studies by \cite{ann62} and \cite{ann66}. In the article~\cite{ann62}, a novel probabilistic approach is presented for imputation in ANNs. Rather than employing single imputations to complete missing values, this method utilizes probability density functions, such as the Gaussian Mixture Model (GMM), to model uncertainty for each missing attribute. This probabilistic perspective is then incorporated into the neural network's processing, notably in the neuron's response within the first hidden layer.

Complementing this, the study by Chen et al.~\cite{ann66} proposes a deep neural network framework with 15 hidden layers, using a Multilayer Perceptron (MLP) to impute missing data. They then evaluate the imputed dataset using a Support Vector Machine (SVM) classifier. Remarkably, their method achieves an accuracy of 89\%, on par with using a complete dataset, suggesting that their deep learning imputation method minimally introduces bias when handling missing values.

The second approach, focusing on training networks with incomplete datasets, is detailed in study \cite{annmlp}. This method diverges from traditional imputation techniques. Instead of attempting to fill in missing data, it proposes robust training strategies that adapt to missing datasets. The process involves dividing the dataset into subsets that are complete in certain clusters of features. An ensemble of base networks is then trained using these subsets. The final step involves merging these base networks and fine-tuning the combined model. This innovative approach allows for effective training of neural networks on datasets that contain only incomplete samples, without the prerequisite of fully observable data during the training phase.

Both approaches offer significant advancements in handling missing data within the context of ANN training and deployment. However, it is important to note that neither method specifically addresses the different mechanisms of missing data, which may limit their suitability for handling certain types of missing data scenarios.

\subsubsection{Flow Based}

Flow-based generative models are powerful machine-learning models for complex data distributions. These models are constructed through a sequence of invertible transformations, making them particularly useful for various tasks such as image, audio, video, and sequence generation~\cite{flowaudio,flowvideo,flowspeech}. The central concept behind flow-based generative models is to transform a simple probability distribution, often a Gaussian distribution, into a more complex distribution that closely resembles the data distribution. This transformation is achieved using invertible transformations, or flows, which efficiently map data from the simple distribution to the complex distribution. It can be computed in both forward and inverse directions.\\\\In recent works, flow-based generative models have demonstrated promising results in generating high-quality samples and addressing missing data imputation tasks~\cite{EMflow,mcflow}. MCFlow~\cite{mcflow} is a novel deep framework explicitly designed for data imputation. It addresses missing data challenges by leveraging normalizing flow generative models and Monte Carlo sampling. The framework employs an iterative learning approach, alternately updating the density estimate and the missing data values in the training data. This iterative learning approach enables MCFlow to explicitly learn complex high-dimensional data distributions, enhancing its capability to handle missing data. The iterative learning scheme efficiently updates the density estimate based on the completed data with imputed values, and a novel non-iterative approach is used for maximum-likelihood sampling in the latent flow space to find optimal values for the missing data.\\\\Similarly, EMFlow~\cite{EMflow} is another approach that combines an Expectation-Maximization (EM) algorithm with normalizing flow (NF) generative models. This integration allows for accurate imputation in a latent space while leveraging the interpretability and numerical stability of EM and the efficient data sampling and representation power of NF. The iterative learning strategy of EMFlow involves updating the density estimation of completed data and the latent space parameters alternatively, refining the initial naive imputation step until convergence. By deriving the, the method consumes data in batches during parameter updates, resulting in faster convergence compared to competing methods .\\\\Despite their benefits, flow-based imputation methods can be computationally expensive, particularly for large datasets and complex transformations. This computational cost should be considered when applying these methods to real-world data analysis tasks.

\subsubsection{Variational Autoencoder}

The Variational Autoencoder (VAE) is a powerful generative deep learning model that combines variational inference and autoencoders. Its main goal is to learn and generate complex data by mapping it into a lower-dimensional latent space. The encoder part of the model transforms input data into a probabilistic distribution in the latent space, capturing meaningful features. The decoder then uses this information to generate new samples with similar characteristics, making VAEs useful not only for data generation but also for data imputation. VAEs can handle various data types, including images, tabular data, video, audio, and even IoT data, making them versatile and widely used in generation and imputation tasks.~\cite{razavi2019generating,peng2021generating,CTGAN,miwae,notmiwae,yan2021videogpt,lee2018stochastic,brunner2018midi,luo2020mg,GINA,MCVAE,partialVAE,GPVAE}.
\\\\
In Mattei et al 's work ~\cite{miwae}, they introduced MIWAE, a method specifically designed to handle missing data in the context of deep latent variable models, particularly in scenarios where the data is MAR. MIWAE is based on the importance-weighted autoencoder (IWAE)~\cite{IWAE} and aims to maximize a tight lower bound of the log-likelihood of the observed data. Notably, MIWAE does not introduce any additional computational overhead due to missing data when compared to the original IWAE. The researchers developed efficient Monte Carlo techniques for both single and multiple imputation using a VAE trained on an incomplete dataset. They demonstrated the effectiveness of MIWAE by training a convolutional VAE on incomplete static binarizations of the MNIST dataset.  \\\\
Ipsen et al.~\cite{notmiwae} proposed the Not-MIWAE, to address the challenge of MNAR data for numerical, categorical, and image datasets. The authors presented a detailed mathematical formulation and derivation of the model's structure and principles, ensuring clarity for readers. By incorporating an additional prior, the Not-MIWAE extends the MIWAE model to satisfy the MNAR assumption by adding a Bernoulli decoder to learning the knowledge the mask. The study conducted experiments on synthetic and real-world datasets, providing replicable methods for handling MNAR data. The comprehensive experimental setup encompassed tabular data (numerical and categorical) and image data. However, while the Not-MIWAE demonstrated superior performance on MNIST images and tabular datasets compared to existing methods, its applicability to real-world scenarios may be limited. The model relies heavily on the Gaussian distribution assumption, which might not always hold, especially for nominal data. As a result, the Not-MIWAE may not be suitable for handling nominal data that often deviates from the Gaussian assumption observed in real-world scenarios. Additionally, when applying MNAR missing mechanisms, the missing rule used in the experiments was relatively simple and not fully representative of real-world scenarios. Despite these limitations, the Not-MIWAE showcased promising results on the tested datasets, providing valuable insights into imputation performance for different data types.\\\\
GINA, introduced by Ma and Zhang~\cite{GINA}, addresses the challenges MNAR data poses in real-world datasets. While a few methods have considered the MNAR scenario, their model's identifiability under MNAR is not guaranteed. The identifiability issue means that model parameters cannot be uniquely determined even with infinite data samples, further contributing to potential bias in the imputation process. Modern deep generative models mainly overlook the issue of identifiability. In their study, they bridge the gap by systematically analyzing the identifiability of generative models under MNAR. Moreover, they propose a practical deep generative model that can provide identifiability guarantees under mild assumptions, accommodating a wide range of MNAR mechanisms.\\
Overall, researchers put lots of effort into special missing mechanisms in VAE based methods. However, most of them are still focusing on a single missing mechanism, and no method can deal with all special mechanisms.

\subsubsection{Generative Adversarial Networks}
Generative Adversarial Network (GAN) was originally proposed by Goodfellow et al.~\cite{goodfellow2014generative}. It is an innovative type of deep learning model. It consists of two neural networks: the generator and the discriminator. The generator creates new data, like images or text, that resembles real data seen during training. On the other hand, the discriminator acts as a "critic" and tries to distinguish between real and generated data. During training, these two networks play in a competitive game. The generator aims to produce data so realistically that the discriminator cannot tell it apart from real data. The discriminator's goal is to detect the generator's fake data better. Similar to the VAE model, the GAN model also has lots of applications for the generation and imputation of various data types, e.g., image generation, time-series data generation, video paragraph generation~\cite{gan1,gan2,gan3,gan4,CTGAN,gain,lee2019collagan,luo2018multivariate,misgan}. \\\\
Yoon et al.~\cite{gain} introduced a novel approach for imputing missing data using the GAN framework, Generative Adversarial Imputation Nets (GAIN). In GAIN, a generator (G) leverages observed components of an actual data vector to impute the missing components and produce a completed vector. A discriminator (D) is employed to distinguish between observed and imputed components, guided by a hint vector that provides additional information about the missingness pattern. The hint helps focus D's attention on the imputation quality of specific components, ensuring that G learns to generate data that adheres to the true data distribution.
\\\\GAMIN~\cite{gamin} is a GAN-based multiple-imputation method proposaed by Yoon. It is designed to handle highly missing data with missing rates exceeding 80\%. While existing imputation methods primarily focus on moderate missing rates, GAMIN also utilizes a GAN framework to generate imputations and incorporates a confidence prediction method to ensure reliable multiple imputations. The proposed method outperforms baseline approaches in experiments conducted on MNIST and CelebA datasets with high missing rates, demonstrating its effectiveness in addressing the challenging problem of highly missing data. GAMIN's innovative imputation architecture, confidence prediction, and specialized learning and inference techniques make it a valuable contribution to the field of imputation for challenging datasets.\\\\Li et al.~\cite{misgan} introduced MisGAN, a GAN-based framework designed for learning from high-dimensional incomplete data. Their method integrates a completed data generator and a mask generator that models the missing data distribution. An auxiliary GAN is utilized to learn a mask distribution that captures the missingness pattern, enabling the 'masking' of the generated data. The completed data generator is subsequently trained to produce masked data that is indistinguishable from real incomplete data. Distinct from prior methods, their framework does not require prior knowledge of the measurement process, making it suitable for a broad range of missing data challenges. Empirical results showcased the capability of their approach in learning intricate data distributions from significantly incomplete data and in generating high-quality imputations. This positions it as a promising solution for addressing missing data in real-world contexts.
The GAN-based imputation methods mentioned above are undoubtedly strong and powerful for handling missing data. However, they overlook the importance of explicitly considering the missing mechanism. Special missing mechanisms require understanding the specific relationships between the observed and missing parts of tabular data, which may not be straightforward for GANs because they do not rely on explicit mathematical derivations. As a result, they may not fully explore and capture the underlying relationships between the observed and missing data, potentially leading to suboptimal imputations. To address this limitation, future research could explore incorporating additional mechanisms or constraints into GAN-based imputation frameworks to enhance their ability to handle missing data with different missing mechanisms effectively.

\subsubsection{Diffusion Model}
Diffusion models have garnered significant attention for their remarkable success in generative modeling across diverse domains, including images, texts, audio, and multi-modality tasks~\cite{diffuse1,csdi,tabcsdi,poole2022dreamfusion,chen2021wavegrad,tevet2022human}. However, their potential to address missing value imputation for tabular data remains relatively unexplored. Recent advancements in score-based diffusion models have demonstrated exceptional performance in tasks like image generation and audio synthesis, sparking interest in their potential for time series imputation. \\\\Conditional Score-based Diffusion models for Imputation (CSDI)~\cite{csdi}, proposed by Tashiro et al., represent a score-based diffusion model tailored for time series data imputation. Leveraging observed data, CSDI effectively imputes missing values using score-based diffusion models. Through extensive evaluation of healthcare and environmental datasets, it exhibits substantial superiority over existing probabilistic imputation methods, achieving remarkable performance improvements of up to 40-65\% on RMSE metrics. CSDI extends beyond time series data and finds applicability in other imputation tasks, such as interpolation and probabilistic forecasting while maintaining competitiveness with existing baselines. However, a limitation of CSDI lies in its suitability for mix-type tabular datasets, which comprise both categorical and numerical variables. \\\\Addressing this gap, Zheng et al.~\cite{tabcsdi} propose TabCSDI, building upon CSDI and exploring three techniques for effectively handling categorical and numerical variables together: one-hot encoding, analog bits encoding, and feature tokenization. Empirical evaluations on benchmark datasets reveal the efficacy of TabCSDI compared to existing methods, emphasizing the importance of employing appropriate categorical embedding techniques in tabular data imputation. It is worth noting that while CSDI and TabCSDI excel in various imputation scenarios, they do not explicitly address specific missing mechanisms, presenting an area for further exploration and improvement.

\subsection{Optimization Algorithm Imputation}
\label{sec:Optimization}
Optimization algorithms are valuable tools in missing data imputation, as they can enhance existing imputation methods and improve their performance. Rather than acting as standalone imputation frameworks, optimization algorithms work alongside imputers to fine-tune their results. These algorithms can optimize the imputation model or refine the imputed dataset, leading to more accurate and effective handling of missing values. The imputation process becomes more robust by leveraging optimization techniques, yielding better-imputed data and enabling more reliable analyses in various applications.

\subsubsection{Data Enhanced}
Genetic Algorithm (GA) is a Data Enhanced optimization technique inspired by biological evolution~\cite{GA74,GA76}. It operates by initializing a population of potential solutions represented as strings (chromosomes) and then applying selection, mutation, and crossover operations, mimicking natural selection. The fittest solutions are chosen, combined, and iteratively refined to create new generations of solutions until a satisfactory outcome is achieved or a predefined number of generations is reached. For example, Shahzad et al. ~\cite{GA80} proposed a novel imputation approach that integrates GA with Information Gain (IG). GA is used to generate optimal sets of missing values in the dataset, while IG is employed to evaluate the performance of each imputation solution. This method is particularly effective for datasets with large search spaces and a higher rate of missing values, enhancing the dataset selection process. The GA imputation method has been extensively explored in the literature to address all three types of missing data mechanisms and various types of variables~\cite{GA80,GA81,GA82}.

\subsubsection{Training Enhanced}
The optimal transport (OT)~\cite{optimaltransport} method is a Training Enhanced imputation approach that modifies the loss function during training. It ensures that two randomly extracted batches from the dataset share the same distribution, making it a suitable loss function for imputing missing values. Extensive experiments demonstrate that the OT-based method outperforms state-of-the-art imputation techniques, even with high percentages of missing values by minimizing these losses through end-to-end learning. OT's ability to provide meaningful distances for comparing distributions makes it well-suited for missing data imputation. The paper presents two algorithms, one non-parametric and one parametric, for imputing missing values based on this loss function. OT's flexibility allows it to combine with simple imputers like MICE~\cite{MICE} and other iterative imputation methods, effectively enhancing their performance in handling missing data across various missing mechanisms.

\subsubsection{Hyperparameters Enhanced}
HyperImputer~\cite{jarrett2022hyperimpute} is a Hyperparameters Enhanced optimization technique, it is an advanced imputation method that combines the benefits of traditional iterative approaches and deep generative modeling to handle missing values in datasets. It automatically configures column-wise models and their hyperparameters, allowing for adaptability across different types of variables and datasets. This approach simplifies the imputation process and reduces the need for manual parameter tuning. Empirical studies have demonstrated that the HyperImputer produces accurate imputations, outperforming existing benchmarks and highlighting the importance of well-configured conditionals for achieving state-of-the-art performance. However, it's important to note that the HyperImputer is most suitable for addressing MAR data, where the probability of missingness depends on observed data but not on the missing value itself. While the method excels in handling MAR missing data mechanisms, it may not be as effective for handling MNAR data, where the missingness is related to unobserved data, posing challenges in imputation. \\\\Table \ref{DLtable} shows the summary of deep learning based and Optimization Based method. Those methods are preferred for missing data imputation due to their inherent advantages. Firstly, their robustness stems from their ability to capture intricate relationships and patterns within data, enabling accurate imputations even in the presence of complex missing mechanisms. Secondly, these methods are efficient, allowing for fast computations and training on large datasets, which is essential for real-world applications. Thirdly, their adaptability across diverse data types, including numerical, categorical, and images. Later on they can be extend to more data types.

\begin{table}[!h]
\centering
\begin{tabular}{|c|c|c|c|c|c|c|}

\hline
     Method&Citation&MCAR&MAR&MNAR&DataType&Evaluation  \\\hline\hline

    

 \multirow{2}*{ANN}& ANN\cite{ann66}& \cmark&\xmark&\xmark & Numerical & ACC\\\cline{2-7}
~& GapNet\cite{annmlp}&  \cmark&\xmark&\xmark & Numerical& AUC\\\hline

 \multirow{2}*{Flow}& MCFlow \cite{mcflow} & \cmark&\xmark&\xmark& Numerical, Image & RMSE,FID,ACC\\\cline{2-7}
~&  EMFlow \cite{EMflow} & \cmark&\cmark&\xmark & Numerical,Image &RMSE,ACC\\\hline

 \multirow{7}*{VAE}& GP-VAE\cite{GPVAE} & \cmark&\xmark&\cmark & Temporal & RMSE, AUC\\\cline{2-7}
    & MCD-VAE\cite{MCVAE} & \cmark&\xmark&\xmark & Numerical & RMSE, ACC\\\cline{2-7}
   &  MIWAE\cite{miwae} &  \cmark&\cmark&\xmark & Numerical , Image  & RMSE,ACC \\\cline{2-7}
   &  Not-MIWAE\cite{notmiwae} &  \cmark&\xmark&\cmark & Numerical, Image  & RMSE,MSE \\\cline{2-7}
   &  GINA\cite{GINA} & \cmark&\xmark&\cmark & Numerical &  Vis$^2$, MSE\\\cline{2-7}
   &  PartialVAE\cite{partialVAE} & \cmark&\xmark&\cmark & Mixed &  MAE\\\hline

 \multirow{4}*{GAN}& GAIN\cite{gain} & \cmark&\xmark&\xmark & Numerical  & RMSE, AUC\\\cline{2-7}
   &  MisGAN\cite{misgan} & \cmark&\cmark&\xmark  & Image &RMSE,FID,ACC \\\cline{2-7}
  &   GAMIN\cite{gamin} & \cmark&\xmark&\xmark & Image & RMSE \\\cline{2-7}
   &  GI\cite{GI} &  \cmark&\xmark&\xmark & Numerical, Image & FID, RMSE\\\hline

     \multirow{2}*{Diffusion}   & CSDI\cite{csdi}&  \cmark&\xmark&\xmark & Temporal& MAE\\\cline{2-7}
    & TabCSDI\cite{tabcsdi}&  \cmark&\xmark&\xmark & Mixed& RMSE\\\hline
    \multirow{3}*{Optimization Based} & GA\cite{GA80}& \cmark & \cmark &\xmark&Numerical & ACC,F1,AUC  \\\cline{2-7}
    &  OT\cite{optimaltransport}& \cmark & \cmark &\cmark&Numerical & RMSE,MAE  \\\cline{2-7}
   & HyperImpute\cite{jarrett2022hyperimpute}& \cmark & \cmark &\xmark&Mixed & RMSE,WD  \\\hline 
\end{tabular}
    
\caption{Deep learning-based and optimization based method Summary\\Vis$^2$: Visualization Plots} 
\label{DLtable}
\end{table}


\section{Representation Learning for Handling Missing Data}
\label{sec:Representation Learning for Handling Missing Data}
Representation learning involves extracting meaningful features from raw data by uncovering its underlying structure and patterns. These learned representations enhance the effectiveness of subsequent tasks. In tasks involving missing values, the focus is not merely on filling in the gaps but leveraging the learned representations to address downstream challenges using incomplete data directly. Furthermore, these representations can be combined with other imputation techniques to enhance performance. Graph Neural Networks (GNNs) and AutoEncoders (AEs) are two common types of representation learning methods. GNNs are often preferred for multi-modality imputation, while AEs are effective for handling complex datasets such as speech, video, and temporal.

\subsection{Graph Neural Networks}
GRAPE~\cite{representationlearning_GNN} stands as a representation learning method designed to address the challenges of missing data. GRAPE uses a label prediction approach, unlike existing methods that often make strong assumptions. This approach aims to directly accomplish downstream tasks, such as classification or regression while dealing with missing values in the input data. GRAPE's uniqueness lies in its adoption of a graph-based representation paradigm. It visualizes observations and features as nodes within a bipartite graph, where edges represent observed feature values. This inventive framework characterizes feature imputation as an edge-level prediction task and label prediction as a node-level prediction task. Using Graph Neural Networks (GNNs), GRAPE introduces several architectural innovations. It incorporates edge embeddings and augmented node features during message passing, effectively enhancing its representation power. Additionally, an edge dropout technique is employed to mitigate overfitting challenges, resulting in an overall enhanced performance of the GRAPE method.\\\\
Malone et al.\cite{representationlearning_GNN_ehr} addressed the challenge of missing data in Electronic Health Records (EHRs) using Graph Neural Networks (GNN). Given the frequent presence of missing data among patients, the diverse data modalities, and the intricate task of identifying crucial patient relationships, the study introduces a representation learning approach based on message passing. This approach consistently outperforms or competes favorably with existing methods in predicting critical medical outcomes such as in-hospital mortality, duration of hospital visits, and discharge destinations. To tackle missing data, the authors extend the embedding propagation framework. They introduce a dual representation for each patient: the initial representation encodes information from observed data, while the second representation is tailored to accommodate missing data. Integrating a dedicated missing data representation within the graph-based learning framework provides the advantage of propagating these representations across the graph, influenced by neighboring patients' representations. Merging these feature representations results in a comprehensive representation that substantially benefits downstream tasks.\\\\
The article~\cite{representationlearning_gnn_spatio} proposes a novel approach to modeling the missing  multivariate time series as temporal signals over dynamic graphs. The method involves autoregressive graph neural networks for recursive learning of representations at each time point and space. They introduce an attention-based architecture that exploits spatiotemporal propagation aligned with the imputation task, effectively reconstructing missing data points. The architecture leverages an inter-node sparse spatio-temporal attention mechanism within the neural message-passing framework, ensuring accurate reconstruction while accounting for missing data.

\subsection{AutoEncoder}
Autoencoders represent another approach to representation learning~\cite{representationlearning_AE_imputation,representationlearning_autoencoder,representationlearning_autoencoder_video}. They comprise an encoder, which compresses input data into a lower-dimensional space and a decoder, which reconstructs the original data. Autoencoders capture essential features while reducing noise by minimizing the discrepancy between input and reconstructed output during training.\\
In the case of TKAE~\cite{representationlearning_autoencoder}, the focus is on learning compressed representations of multivariate time series (MTS) data, which enhances data analysis by addressing noise, redundancy, and handling numerous variables and time steps. Conventional dimensionality reduction methods are ill-suited for MTS due to their inability to manage missing values. Researchers propose a novel autoencoder architecture based on recurrent neural networks to overcome this limitation. This autoencoder is tailored for MTS, accommodating variable lengths and effectively handling missing data. The model learns fixed-length vector representations, aligning their pairwise similarities using a kernel function that operates in the input space and manages missing values. The efficacy of this approach is validated through classification tasks, including medical data, and compared against other dimension reduction techniques.

\begin{table}[!h]
\centering
\begin{tabular}{|c|c|c|c|c|c|c|}

\hline
     Method&Citation&MCAR&MAR&MNAR&DataType&Evaluation  \\\hline\hline

    

 \multirow{2}*{GNN}& GRAPE\cite{representationlearning_GNN}& \cmark&\xmark&\xmark & Tabular & MAE\\\cline{2-7}
~& SpatioGNN\cite{representationlearning_gnn_spatio}& \cmark&\xmark&\xmark & SpatioTemporal & MAE\\\cline{2-7}
~& EHRGNN\cite{representationlearning_GNN_ehr}&  \cmark&\xmark&\xmark & Multimodality, Temporal& MAE\\\hline

 \multirow{1}*{VAE}& TKVE \cite{representationlearning_autoencoder} & \cmark&\xmark&\xmark& Temporal & AUC,MSE\\\hline

\end{tabular}
    
\caption{Representation Learning based method Summary} 
\label{RLtable}
\end{table}


\section{Existing Methodology for Missing data Generation}
\label{sec:Methodology for Missing data Generation}
Researchers often face the challenge of dealing with special missing mechnisam patterns in real data, where most of statistical methodologies are unavailable to identify them. In such cases, researchers heavily rely on their judgment and domain knowledge. For experimental studies, researchers commonly manually simulate MNAR and MAR data to investigate its characteristics. In this section, we explore common MNAR and MAR generation methods for tabular data, providing valuable insights for researchers dealing with missing data.

\subsection{MAR Generation Methods}
Recalling the definition from section \ref{sec:MAR}, the Missing At Random (MAR) mechanism arises in cases where the missing portion $X^m$ is dependent on the observed portion $X^o$. This relationship can be represented by the following formula:
Two distinct subtypes of the MAR Generation Method exist:
$$
f(\boldsymbol{M} | X^o, \Psi) \textbf{  }  \forall \textbf{  } X^m,\Psi
$$
There are two subtypes of MAR Generation Method:
\begin{itemize}
    \item Threshold Method\\
    The generation of MAR data involves a threshold, defined based on $X^o$, along with a rule. For instance, by utilizing the column mean of $X^o$ as the threshold and a specific rule, any value exceeding the threshold is excluded from $X^m$.
    \item Probability Method\\
This method employs a probability function to determine the missing probabilities. Ordinarily, the observed portion $X^o$ is fed into the probability function, generating a probability value that dictates whether the data $x_i^m$ will be missing. Commonly used functions for this purpose include the \textit{sigmoid} function~\cite{mcflow, miwae} and the \textit{Logistic} model ~\cite{misgan, jarrett2022hyperimpute}.
    
\end{itemize}

\subsection{MNAR Generation Methods}
The Missing Not At Random (MNAR) mechanism is the most intricate among the various missing mechanisms. Sections \ref{sec:MNAR} introduced the concept of MNAR and its subtypes. Numerous scholarly works have explored diverse approaches to generate MNAR data. Gomer and Yuan~\cite{Subtypesofthemissing} offer a comprehensive compilation of studies on MNAR data generation.
\subsubsection{Focused MNAR Generation Method}
\begin{figure}[!h]
    \centering
    \includegraphics[scale = 0.5]{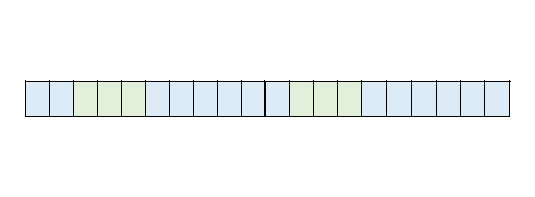}
    \caption{Brusty Missing Example, blue means present data, green means missing data }
    \label{fig:Brusty}
\end{figure}

Within the realm of MNAR generation methods, several subtypes have been identified, each with distinct characteristics:
\begin{itemize}
\item Quantile Method\\
The Quantile Method~\cite{zhang2012note, notmiwae} generates missing values based on the percentiles of $X^m$. Specifically, $X^m$ values above or below the cth percentile of $X^m$ are more likely to exhibit missingness.

\item Threshold Method~\cite{Subtypesofthemissing}\\This method generates missing values for variable $X^m$ based on a predetermined threshold. An auxiliary variable $Z$, correlated with $X^m$, is employed. When $Z$ surpasses the threshold, the likelihood of missingness in $X^m$ values increases.

\item Correlated Auxiliary Variable~\cite{zhang2012note}\\ Missing values within $X^m$ are generated based on the correlation between $X^m$ and an auxiliary variable $Z$. the probability of missing values on $X^m$ increases with larger $Z$ values. 

\item Probability Method~\cite{optimaltransport}\\Similar to the MAR probability method, the Probability Method for MNAR requires $X^m$ itself to be fed into the probability model for obtaining probability values. Notably, methods like MIWAE\cite{miwae} and HyperImputer utilize a \textit{logistic} model for this purpose.

\item Bursty Method ~\cite{dlreview}\\This method pertains to instances where missing data appear to cluster within specific groups, orders, or temporal moments. Refer to Figure \ref{fig:Brusty} for an illustrative example. In this approach, a corresponding number of bursts are randomly selected to invalidate data, simulating the desired impairment level within the dataset. For instance, when the dataset size is 20 and a 30\% impairment rate is desired, with a burst size of 3 (resulting in 6 data cells invalidated across 2 bursts of size 3), the dataset is manipulated accordingly.
\end{itemize}

\begin{figure}[!h]
    \centering
    \includegraphics[scale = 0.5]{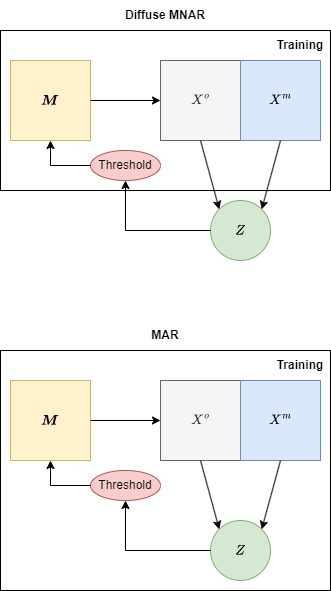}
    \caption{Diffused MNAR and MAR for  Correlated Auxiliary}
    \label{fig:Diffused MNAR and MAR for  Correlated Auxiliary}
\end{figure}


\subsubsection{Diffuse MNAR Generation Method}
This section highlights the Diffuse Missing Not At Random (MNAR) generation methods, which encompass the following subtypes:
\begin{itemize}
    \item Linear Combination with Variable $Z$ ~\cite{zhang2012note}\\ In this approach, $Z$ is generated through a linear combination of $X^o$ and $X^m$. Missingness in $X^m$ is contingent upon $Z$ surpassing a predefined threshold.
    
    \item Correlation with Auxiliary Variable $Z$~\cite{collins2001comparison}\\This method defines the missingness of $X^m$ based on the correlation between an auxiliary variable $Z$ and $X^o$. Additionally, $Z$ and $X^m$ share a correlation. The procedure involves partitioning the data into groups and calculating the sample correlation between $X^o$ and $Z$ within each group. $X^m$ values from groups with high correlations are likelier to exhibit missingness than those from low-correlation groups.

\end{itemize}
For both methods, excluding $Z$ results in diffuse MNAR. When $Z$ is omitted from the analysis, the generated missing values also follow the diffuse MNAR pattern. In contrast, if $Z$ is incorporated, the missingness conforms to MAR standards. Refer to Figure \ref{fig:Diffused MNAR and MAR for Correlated Auxiliary} for visual representation.

The preceding sections have introduced a lot of generation methods that have been employed in various experiments. However, the practical implementation of these methods can significantly diverge based on the chosen parameters for the missing mechanism ($\Psi$), the involvement of auxiliary variables ($Z$), and the division between $X^m$ and $X^o$. These variations in implementation contribute to differences in the resultant missing data and subsequent imputation procedures.

\section{Experiments and Evaluation Metrics}
In this section, we provide an overview of the standard experimental process and commonly employed evaluation metrics for the missing data imputation task.
\label{sec:Evaluation Metrics}

\subsection{Experimental Process}
\begin{enumerate}
    \item \textbf{Dataset Selection:}
    The experimental process commences with the selection of diverse real-world datasets from various domains, including healthcare, sensor data, images, and synthetic datasets. For tasks spanning different fields, researchers often utilize publicly available datasets. Tabular data can be sourced from the UCI Machine Learning Repository\footnote{https://archive.ics.uci.edu/}, while image datasets such as MNIST\footnote{http://yann.lecun.com/exdb/mnist/}, CelebA\footnote{https://mmlab.ie.cuhk.edu.hk/projects/CelebA.html}, and CIFAR-10\footnote{https://www.cs.toronto.edu/~kriz/cifar.html} are common choices.

    \item \textbf{Missing Value Generation and Data Split:}
    As introduced in Section \ref{sec:Background and Preliminary}, the three missing data parameters - missing rate, missing pattern, and missing mechanism - are crucial considerations. Researchers must apply these missing strategies to the chosen datasets and utilize a mask $\boldsymbol{M}$ to track the locations of missing values. The datasets are subsequently split into training, validation, and test sets, while maintaining corresponding masks for evaluation purposes. It's important to note that scenarios involving multiple missing strategies within a single dataset or differing strategies are applied between training and test sets are contingent on the research question.

    \item \textbf{Dataset Standardization:}
    Applying standardization techniques ensures uniform variable scaling and mitigates bias. Common normalization methods include Min-Max Scaling and Standard Scaling. When employing scalers, it is essential to use the appropriate mask to account for missing values. This prevents the scaler from gaining prior knowledge of the dataset, which could lead to misleading results.

    \item \textbf{Training and Parameter Tuning:}
    Training processes vary based on the model architecture, with training sets used for model training and validation sets employed for parameter tuning.

    \item \textbf{Evaluation:}
    The model's utility is tested using the test set. Evaluation metrics depend on the specific research task and will be detailed in the subsequent section.

\end{enumerate}

\subsection{Evaluation Metrics}
\subsubsection{Visualization}

\begin{figure}[!h]
    \centering
    \includegraphics[width=1\linewidth]{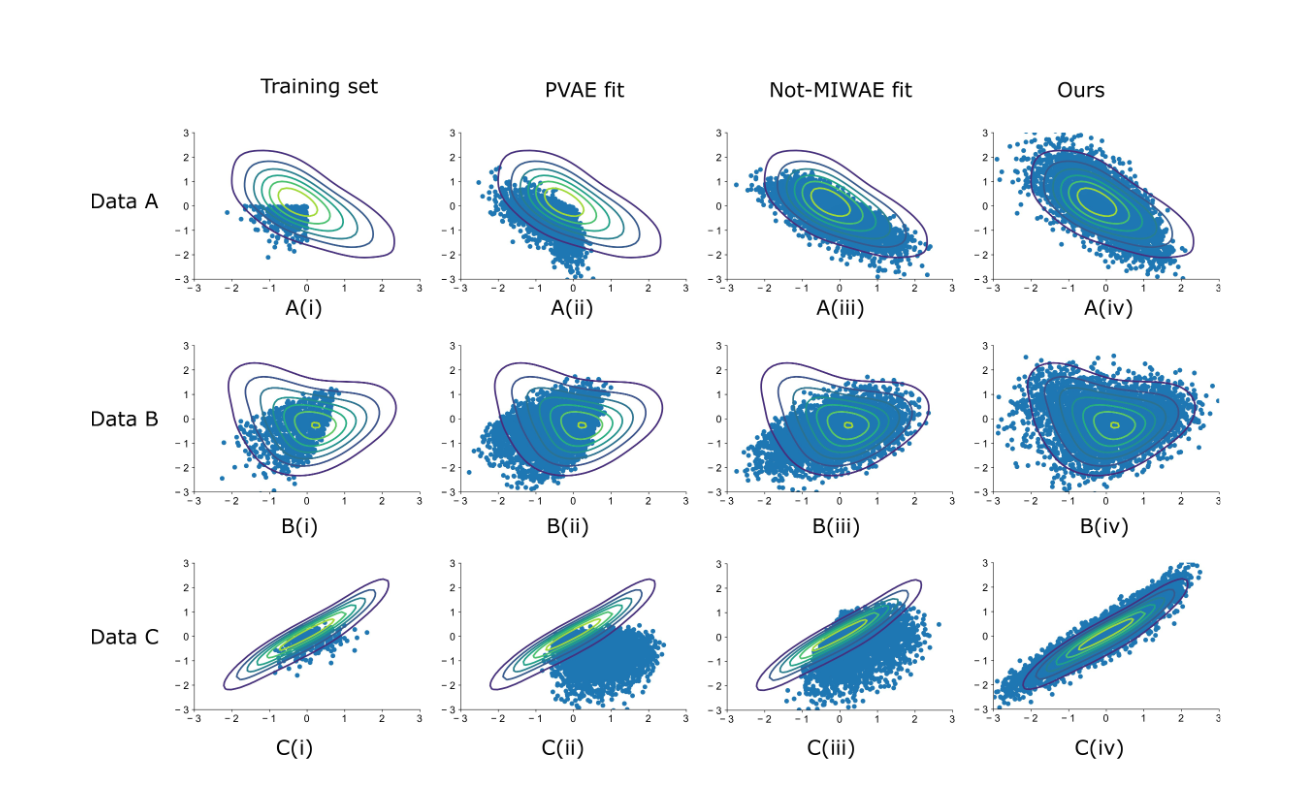}
    \caption{Scatter Plot and Contour Plot from GINA\cite{GINA}}
    \label{fig:Contour}
\end{figure}

\begin{figure}[!h]
    \centering
    \includegraphics[width=0.8\linewidth]{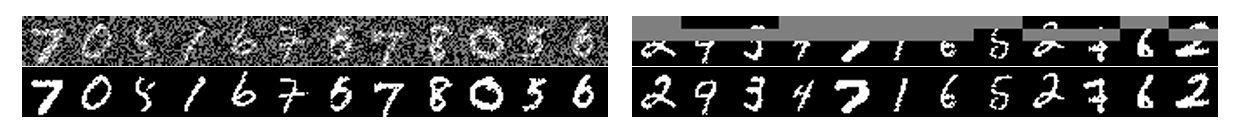}
    \caption{MNIST from~\cite{miwae}}
    \label{fig:MNIST}
\end{figure}

Visualization plays a crucial role in evaluating the quality of imputed missing data. Scatter plots are commonly used to compare imputed values against true values, visually depicting them as points with distinct colors. A reliable imputation method should exhibit a similar pattern to the original data, as shown in Figure \ref{fig:Contour} where scatter and contour plots are utilized for GINA~\cite{GINA} to present a 2D synthetic dataset, facilitating easy comparison of different imputers. However, scatter plots are most effective for low-dimensional data or data with strong internal correlations. In contrast, high-dimensional data demands alternative visualization techniques to effectively assess imputation performance. In the case of image datasets like the incomplete MNIST data and its imputation result depicted in Figure \ref{fig:MNIST}, visualization is particularly useful for examining imputer quality. However, the focus has shifted towards tabular datasets, making plotting resulting images less necessary. Instead, column density plots offer a viable option. Nonetheless, it is worth noting that complex missing mechanisms may require a comprehensive analysis of the mix column effect rather than solely concentrating on individual columns. As most existing method are working on high dimensional tabular data, thus visualizations are usually not the best choice for evaluation.

\subsubsection{Quantitative Analysis}
Quantitative analysis involves the use of distance-based methods to evaluate the performance of imputation techniques compared to the original data. These distance measures quantify the differences between imputed values and their corresponding true values.. To ensure unbiased evaluation, the distance is calculated only for the missing values and their imputed counterparts.
\begin{itemize}
    \item \textbf{Root Mean Squared Error (RMSE)}\\
RMSE is widely employed as an evaluation metric in many studies. It is preferred due to its ability to penalize larger errors more than smaller ones, allowing for a more accurate assessment of imputation quality. The formula for RMSE is given by:
$$
\text{Root Mean Squared Error } = \sqrt{\frac{1}{n_{m}}\sum^{n_m}_{i}(X_i^{m} - \bar{X}_i^{m} )^2}
$$
where $X_i^{m}$ represents the true value of the missing part, 
$\bar{X}_i^{m}$  denotes the imputed value, and 
$n_m$  indicates the number of missing values. A lower RMSE value is preferred, indicating better imputation performance. However, it is essential to consider that RMSE is sensitive to outliers in the data. Therefore, researchers often complement RMSE with other evaluation metrics to gain a more comprehensive understanding of the imputation method's effectiveness.
\item \textbf{Mean Squared Error (MSE)}\\
MSE is also a commonly used evaluation metric,
$$
\text{Mean Squared Error } = \frac{1}{n_{m}}\sum^{n_m}_{i}(X_i^{m} - \bar{X}_i^{m} )^2
$$
where $X_i^{m}$ represents the true value of the missing part, 
$\bar{X}_i^{m}$  denotes the imputed value, and 
$n_m$.
MSE has a higher unit order than the error unit, this is due to the square of the error.
\item \textbf{Mean absolute Error (MAE)}\\
The MAE is also considered one of the most intuitive evaluation metrics as it calculates the average absolute difference between the actual and imputed values:
$$
\text{Mean Squared Error } = \frac{1}{n_{m}}\sum^{n_m}_{i}|X_i^{m} - \bar{X}_i^{m} |
$$
MAE is preferred for its simplicity and straightforward interpretation. By using the absolute value, it does not distinguish between over-performance or under-performance of the imputation model, making it more robust to extreme errors. This property is valuable when assessing imputation accuracy, especially in situations where large errors may not be disproportionately penalized. MAE provides a more balanced and easily interpretable measure of imputation performance. Researchers often complement MAE with other evaluation metrics like RMSE to gain a more comprehensive understanding of the strengths and weaknesses of the imputation method.
\item \textbf{Fréchet inception distance(FID)}\\
FID is a metric used to assess the quality of generated images from generative models, particularly within the field of image synthesis. This metric offers a quantitative measurement of the similarity between the distribution of real images and the distribution of generated images created by a generative model. Notably, in scenarios where the imputation task involves images,\cite{mcflow,misgan,GI} researchers often employ FID to gauge the effectiveness of imputed images.
$$
\text{FID}(P_{\text{real}}, P_{\text{generated}}) = \|\mu_{\text{real}} - \mu_{\text{generated}}\|^2 + \text{Tr}\left(\Sigma_{\text{real}} + \Sigma_{\text{generated}} - 2\sqrt{\Sigma_{\text{real}}\Sigma_{\text{generated}}}\right)
$$
Where $P$ represents the distribution, $\mu$ is the  means data feature representations, $\Sigma$ is the covariance matrices of feature representations, $\text{Tr}$ denotes the trace of a matrix.

\item \textbf{Log-likelihood} \\ 
Log-likelihood is commonly used as an evaluation metric for generative models that are probabilistic in nature. These models aim to learn the underlying probability distribution of the real data and then generate new data points from that distribution. Some of probabilistic generative models that often use log-likelihood as an evaluation metric Eg.VAE,Probabilistic Matrix Factorization and  Probabilistic Principal Component Analysis~\cite{notmiwae,hernandez2014probabilistic}.

\end{itemize}

\subsubsection{Downstream Tasks}
Evaluating the utility of imputed data is a crucial step in assessing the effectiveness of an imputation method. Researchers often compare the performance of specific tasks using imputed values and the original complete data to identify any differences. This approach not only helps to understand the dataset's underlying distribution but also reveals hidden latent spaces. By evaluating the imputed data's performance in various downstream tasks, such as machine learning models or predictive tasks, researchers gain insights into the imputation method's effectiveness and usefulness. This combined approach allows for a comprehensive evaluation of imputed data quality and provides a deeper understanding of the dataset's underlying structure and characteristics.\\\\
For real-world applications or scenarios where the imputed data will be used in downstream tasks, researchers employ machine learning models to evaluate the imputation methods. As shown in Figure \ref{fig:DownstreamTasks}, the original complete dataset and the imputed dataset are used to train two machine learning models with exactly the same parameter settings, model architectures, and training sizes. Both models are then applied to the same test dataset, and the performance of the predictions is compared. Ideally, the goal is to ensure that the imputed data's utility and predictive ability are similar to those of the complete dataset. Thus, if the true utility and imputed utility are similar, it indicates that the imputation method works effectively. For instance, in the study by~\cite{ann66}, the performance of the deep learning imputation method was evaluated using a support vector machine classifier on a dataset of 799 youths with ADHD and 421 without ADHD. The results showed that the classifier achieved 89\% accuracy on the imputed dataset, which was comparable to the accuracy obtained on the original dataset without any missing values. This demonstrates the success of the imputation method in preserving the utility of the data for downstream tasks.
\\\\
Researchers also commonly utilize Recommendation Systems as a downstream task to evaluate the effectiveness of imputation methods~\cite{GINA,notmiwae,marlin2009collaborative}. In this approach, the imputation task is treated as a recommendation problem, where the imputed values serve as recommendations to be compared with the actual missing values. To evaluate the recommendations made by the recommendation system, various metrics are employed, such as accuracy (ACC), precision, recall, F1-score, Mean Average Precision (MAP), or AUC. These metrics measure the accuracy and relevance of the imputed values as recommendations. By using a recommendation system for evaluation, researchers can gain insights into how well the imputation method performs in providing meaningful and accurate recommendations for the missing values. This evaluation approach offers a practical and real-world perspective on the imputation method's performance, especially in scenarios where the imputed data will be used for recommendation-based applications or decision-making processes.

\usetikzlibrary{trees,positioning,shapes,shadows,arrows.meta}

\definecolor{g1}{HTML}{eaf2ed}
\definecolor{g2}{HTML}{c5ddcf}
\definecolor{g3}{HTML}{9fc9b1}
\definecolor{g4}{HTML}{79b493}
\definecolor{g5}{HTML}{539f75}
\definecolor{g6}{HTML}{2e8b57}

\definecolor{b1}{HTML}{f0f0f1}
\definecolor{b2}{HTML}{dfe4ec}
\definecolor{b3}{HTML}{cdd7e6}
\definecolor{b4}{HTML}{bccbe1}
\definecolor{b5}{HTML}{aabedc}
\definecolor{b6}{HTML}{99b2d6}
\definecolor{b7}{HTML}{88a5d1}
\definecolor{b8}{HTML}{7799cc}

\definecolor{bb1}{HTML}{c8d5de}
\definecolor{bb2}{HTML}{a1baca}
\definecolor{bb3}{HTML}{7a9fb6}

\definecolor{r1}{HTML}{f1f0f0}
\definecolor{r2}{HTML}{e6d2ce}
\definecolor{r3}{HTML}{dab5ad}
\definecolor{r4}{HTML}{ce978c}
\definecolor{r5}{HTML}{c37a6b}
\definecolor{r6}{HTML}{b75c49}

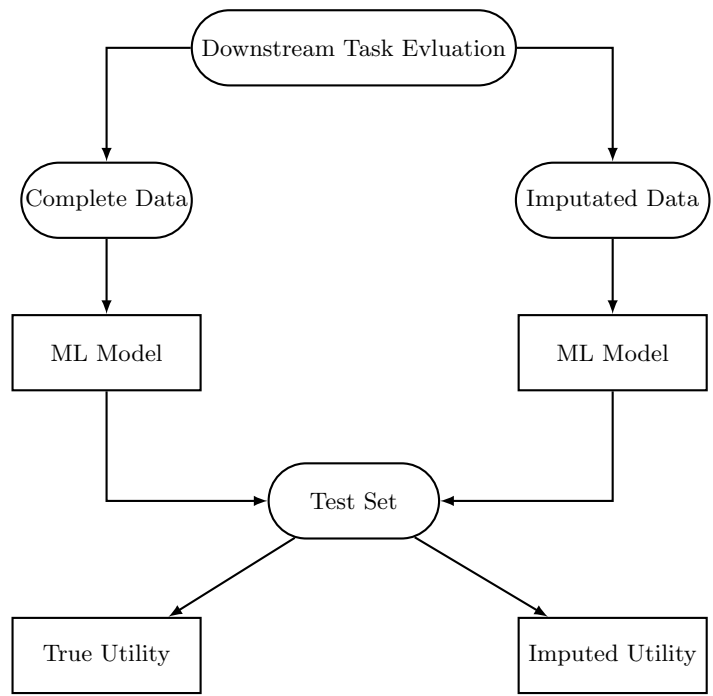
\begin{figure}[!h]
    \centering

\begin{tikzpicture}[font=\small,thick]

\node[draw,
    rounded rectangle,
    minimum width=2.5cm,
    minimum height=1cm] (block4) { Downstream Task Evluation};

\node[draw,
    rounded rectangle,
    below left=of block4,
    minimum width=2.5cm,
    minimum height=1cm] (block5) { Complete Data};

\node[draw,
    rounded rectangle,
    below right=of block4,
    minimum width=2.5cm,
    minimum height=1cm] (block6) { Imputated Data};

\node[draw,
    below=of block5,
    minimum width=2.5cm,
    minimum height=1cm] (block7) {ML Model};


\node[draw,
    align=center,
    below=of block6,
    minimum width=2.5cm,
    minimum height=1cm] (block9) { ML Model};

\node[draw,
    rounded rectangle,
    below=5cm of block4,
    minimum width=2.5cm,
    minimum height=1cm,] (block11) {Test Set};

\node[coordinate,below=4.35cm of block4] (block12) {};

\node[draw,
    align=center,
    below=3cmof block7,
    minimum width=2.5cm,
    minimum height=1cm] (Result1) {True Utility};

\node[draw,
    align=center,
    below=3cmof block9,
    minimum width=2.5cm,
    minimum height=1cm] (Result2) {Imputed Utility};

\draw[-latex] (block4) -| (block5);
    
\draw[-latex] (block4) -| (block6);

\draw[-latex] (block5) -- (block7);
\draw[-latex] (block6) -- (block9);
\draw[-latex] (block7) |- (block11);
\draw[-latex] (block9) |- (block11);

\draw[-latex] (block11) -- (Result1);
\draw[-latex] (block11) -- (Result2);

\end{tikzpicture}

    \caption{Downstream Task Evaluation Flowchart}
    \label{fig:DownstreamTasks}

\end{figure}


\section{Limitation and Future Direction}
\label{sec:Limitations}
Despite the progress and advantages of various imputation methods, there are several limitations that need to be acknowledged. These limitations provide opportunities for future research and improvements in the field of missing data imputation.

\subsection{Limitations}
\begin{itemize}
    \item \textbf{Complex Special Missing Mechanisms:} \\The landscape of special missing mechanisms is characterized by complexity, encompassing a wide range of assumptions and diverse data generation methods. As a result, a standardized and straightforward approach for handling these mechanisms is currently lacking. The intricacies of special missing mechanisms often involve various scenarios and assumptions, leading to a multitude of methods for generating missing data. This diversity presents a challenge in establishing a universally applicable methodology that researchers can follow.\\\\ Current missing data imputation models often exhibit limited consideration for complex special missing mechanisms. Particularly, when dealing with real-world datasets, identifying the specific missing mechanism can be a daunting task, necessitating a deep understanding of the domain. In practice, the complexity of these mechanisms renders them difficult to discern, often requiring domain-specific expertise for effective recognition. As a result, existing models may overlook the nuances of special missing mechanisms, limiting their ability to provide accurate imputations.\\\\ While some researchers have attempted to address special missing mechanisms through imputation techniques, the methods they employ for generating missing data exhibit substantial variability. This lack of uniformity in generating missing data further complicates the comparison of model performance. Each approach may adopt a distinct strategy based on specific assumptions about the missing mechanism, leading to disparate outcomes. Consequently, the comparison of these models' true capabilities becomes challenging due to the inconsistent methodologies employed for generating missing data.

    \item \textbf{Non-Robustness of Traditional Approaches:} \\Traditional machine learning and statistical-based imputation methods can also be applied to address certain special missing mechanisms. However, with the increasing complexity and scale of modern datasets, the efficacy of these methods may diminish, often demanding greater computational resources for accurate imputations. Consequently, the overall efficiency of traditional approaches can decrease, contributing to the declining popularity of these methods in contemporary data analysis. While these methods may retain interpretability, their capability to handle large-scale data is compromised. As a result, a trade-off between interpretability and model utility must be considered when opting for traditional approaches in scenarios involving special missing mechanisms.

    \item \textbf{Limitations to Data Type:} \\
    Certain imputation techniques are inherently tailored to handle numerical data and may encounter difficulties when confronted with categorical or mixed data types. The diverse nature of real-world datasets, encompassing a variety of data formats, exposes the limitations of these methods. While some traditional machine learning and statistical-based imputation models can manage mixed data types, the consideration of mixed data types is relatively scarce in neural network-based models. This is often due to the fact that different methodologies for handling categorical data can significantly impact the model's outcomes. Nonetheless, the importance of addressing special missing mechanisms remains consistent even for binary and ordinal data types, underscoring the need for comprehensive solutions across various data formats.
\end{itemize}

\subsection{Future Directions}
As the field of missing data imputation continues to evolve, several promising directions can be combined to enhance the capabilities and applicability of existing methods. These future directions collectively aim to address the challenges posed by complex missing mechanisms, improve model robustness, extend the scope of imputation to diverse data types and modalities, and leverage domain knowledge for enhanced imputation outcomes. The following points outline key areas for integrated future research:

\begin{itemize}
\item \textbf{Extension to Special Missing Mechanisms:} Special missing mechanisms introduce complexities and assumptions that are diverse and varied. Future research can extend existing imputation methods to handle a broader range of special missing mechanisms, including those based on Generative Adversarial Networks (GANs), flow-based models, and diffusion-based techniques. Ensuring that imputation methods can address a variety of missing mechanisms enhances their utility across different scenarios.

\item \textbf{Addressing Multimechanism Missing Data:} In scenarios involving the coexistence of distinct missing mechanisms, imputation models need to tackle complex combinations of these mechanisms. Future research can explore methods that effectively handle multi-mechanism missing data, providing accurate imputations for real-world datasets with intricate missing patterns.

\item \textbf{Benchmarking and Standardization:} To enable fair comparisons and evaluations of imputation methods, standardized benchmark missing data generation method and common evaluation metrics are pivotal. This approach promotes a comprehensive understanding of method strengths and weaknesses and facilitates consistent assessments of imputation model performance.

\item \textbf{Domain Knowledge Integration:} Incorporating domain knowledge into imputation models offers a practical approach to handle complex missing mechanisms. Use Knowledge Distillation model to do pre-studies or the inclusion of domain-specific information can guide the imputation process, leading to improved outcomes. Exploring methods for seamlessly integrating domain knowledge into imputation models can further enhance their performance.

\end{itemize}

By combining these integrated future directions, researchers can collectively contribute to the advancement of missing data imputation methods, addressing complex missing mechanisms, enhancing robustness, accommodating diverse data types and modalities, and leveraging domain knowledge. This integrated approach provides a holistic perspective on tackling the evolving challenges in the field and promoting more effective and comprehensive imputation solutions.


\bibliographystyle{plain}
\bibliography{sample}

\end{document}